\documentclass[fleqn,usenatbib]{mnras}

\usepackage{newtxtext,newtxmath}

\usepackage[T1]{fontenc}

\DeclareRobustCommand{\VAN}[3]{#2}
\let\VANthebibliography\thebibliography
\def\thebibliography{\DeclareRobustCommand{\VAN}[3]{##3}\VANthebibliography}

\usepackage{graphicx}	%
\usepackage{amsmath}	%

\usepackage{amssymb}	%

\usepackage{xcolor}
\usepackage{gensymb}

\title[Stellar-mass microlensing of GWs]{Stellar-mass microlensing of gravitational waves} %

\author[M. H. Y. Cheung et al.]{
Mark H. Y. Cheung,$^{1}$\thanks{E-mail: hycheung@link.cuhk.edu.hk}
Joseph A. J. Gais,$^{1}$
Otto A. Hannuksela$^{2,3}$
and  Tjonnie G. F. Li$^{1}$
\\
$^{1}$Department of Physics, The Chinese University of Hong Kong, Shatin, N.T., Hong Kong.\\
$^{2}$Nikhef – National Institute for Subatomic Physics, Science Park, 1098 XG Amsterdam, The Netherlands\\
$^{3}$Institute for Gravitational and Subatomic Physics (GRASP), Department of Physics, Utrecht University, Princetonplein 1, 3584 CC Utrecht, The Netherlands
}

\date{\today}

\pubyear{2020}

\begin{document}
\label{firstpage}
\pagerange{\pageref{firstpage}--\pageref{lastpage}}
\maketitle

\begin{abstract}
When gravitational waves pass through the nuclear star clusters of galactic lenses, they may be microlensed by the stars. 
Such microlensing can cause potentially observable beating patterns on the waveform due to waveform superposition and magnify the signal. 
On the one hand, the beating patterns and magnification could lead to the first detection of a microlensed gravitational wave. 
On the other hand, microlensing introduces a systematic error in strong lensing use-cases, such as localization and cosmography studies. 
We show that diffraction effects are important when we consider GWs in the LIGO frequency band lensed by objects with masses $\lesssim 100 \, \rm M_\odot$. 
We also show that the galaxy hosting the microlenses changes the lensing configuration qualitatively, so we cannot treat the microlenses as isolated point mass lenses when strong lensing is involved. 
We find that for stellar lenses with masses $\sim 1 \, \rm M_\odot$, diffraction effects significantly suppress the microlensing magnification. 
Thus, our results suggest that gravitational waves lensed by typical galaxy or galaxy cluster lenses may offer a relatively clean environment to study the lens system, free of contamination by stellar lenses. 
We discuss potential implications for the strong lensing science case. 
More complicated microlensing configurations will require further study.
\end{abstract}


\section{Introduction}

Both gravitational waves (GWs) and electromagnetic (EM) waves can be gravitationally lensed when propagating near massive astrophysical objects, such as galaxies and galaxy clusters~\citep{Ohanian:1974ys,Thorne:1982cv,Deguchi:1986zz,Wang:1996as,Nakamura:1997sw,Takahashi:2003ix}.
To date, only the lensing of EM waves have definitively been observed, but several searches for GW lensing have already been performed~\citep{Hannuksela:2019kle,Li:2019osa,McIsaac:2019use,Pang:2020qow,Dai:2020tpj,Liu:2020par}. 

Suppose the GW is lensed by a galaxy or a galaxy cluster only. 
In that case, we can use the geometrical optics approximation to solve for the trajectory, the arrival time of distinct GW images, and their image types~\cite{Takahashi:2003ix}. 
These GW images will only differ in their amplitudes, arrival times, and overall phases.\footnote{This generally applies to (2,2) modes. 
However, if the GWs exhibit higher-order modes or precession, the so-called saddle point images (type-II) could experience nontrivial waveform changes~\citep{Dai:2017huk,Ezquiaga:2020gdt}.}
Thus, we can identify the strongly lensed images as repeated events with identical frequency evolution using matched filtering and Bayesian analysis~\citep{Haris:2018vmn,Hannuksela:2019kle,Li:2019osa,McIsaac:2019use,Dai:2020tpj,Liu:2020par}.

Several exciting science cases have been proposed for such strongly lensed events. 
For example, suppose we observe four GW images of a single event. In that case, it will allow us to uniquely match the GW image properties to any given lensed galaxy, which may enable us to localize the host galaxy and pinpoint the location of the merger event within the galaxy~\citep{hannuksela2020localizing}. 
Two images might still allow us to constrain the number of candidates~\citep{Sereno:2011ty,Yu:2020agu}, but to a lesser degree as we will need to rely mainly on the magnification ratios to pinpoint the source location.\footnote{Thus, the image properties are largely degenerate with the lens and source properties~\citep{hannuksela2020localizing}. In particular, we will need to rely on magnification ratio measurements, subject to much more uncertainty than the time delays, to identify the source position. Four images allow us to map the position uniquely based on the time delays alone.} 
However, if a galaxy cluster lenses the event, even two events might allow localization, owing to the rarity of such clusters~\citep{Smith:2017mqu,Smith:2018gle,Smith:2019dis,Robertson:2020mfh,Ryczanowski:2020mlt}. 

If we localized such strongly lensed GWs, we might be able to perform precision cosmography studies at high redshift~\citep{Sereno:2011ty,Liao:2017ioi,Cao:2019kgn,Li:2019rns,hannuksela2020localizing}, 
test the speed of gravity at high precision~\citep{Collett:2016dey, Baker:2016reh,Fan:2016swi}, 
connect binary black holes to their host galaxy properties and investigate formation channels~\citep{Chen:2016tys,hannuksela2020localizing}, 
and conduct improved tests of the GW polarization content~\citep{Goyal:2020bkm}.\footnote{We note that these latter polarization tests could be done even for non-localized events~\citep{Goyal:2020bkm}. Still, the localization would significantly improve the tests, as we could solve the sky location and the image properties independently in the electromagnetic band~\citep{hannuksela2020localizing}. } 
Other intriguing use-cases in, e.g., cosmography, have been documented in~\citep{oguri2019strong}. 
However, most of these science cases, as well as the localization studies, rely on a clean environment devoid of systematic errors introduced due to microlensing by stars.\footnote{For example, a substantial fraction of the error budget in the localization studies proposed in Ref.~\cite{hannuksela2020localizing} is dominated by microlensing. }

In particular, when GWs are strongly lensed by galaxy and galaxy clusters, their trajectories might pass through dense nuclear star clusters, introducing a non-negligible chance of microlensing~\citep{Christian:2018vsi}.
The stellar lens would then introduce additional microlensing effects.
In gravitational-wave lensing localization studies, they can introduce a sizable error in the magnification ratio or even the time-delay measurement when the geometrical optics limit is assumed~\citep{hannuksela2020localizing}. 
In general, microlensing could lead to systematic errors which will be counter-beneficial to studies of lensed transient events (see discussion in~\cite{oguri2019strong}) and indeed almost any study relying on gravitational-wave localization (e.g.,~\cite{Sereno:2011ty,Collett:2016dey,Baker:2016reh,Fan:2016swi,Liao:2017ioi,Cao:2019kgn,Li:2019rns,hannuksela2020localizing}). 

Luckily, as GWs have much longer wavelengths than typical EM waves, diffraction effects can suppress microlensing when the stellar lens has an Einstein radius comparable to the wavelength of the GWs~\citep{oguri2019strong}.
However, %
microlenses in the wave optics limit have been mostly studied in isolation (see, however, \cite{Diego:2019lcd} and \cite{Diego:2019rzc} for an interesting exploration of microlensing effects in the extreme high-magnification regime).

We consider the typical case of a stellar lens on top of a galactic lens model at moderate magnification, similar to the usual strong lensing system we expect to observe.
We focus on the waveform suppression effects brought forth by wave optics diffraction and elaborate on its implications. 
Much of our work is motivated by the discussion in Ref.~\cite{oguri2019strong}, which discusses some of these wave optics suppression effects; our work aims to quantify the wave optics suppression when macromodels are involved and investigate their dependence on the microlens properties and position in order to obtain a qualitative understanding. 

We first lay out the procedures for treating a compound lens system consisting of lenses with masses and length scales orders of magnitudes apart.
Then, we show that the geometrical optics approximation is not sufficient to treat the case of stellar-mass microlensing for LIGO/Virgo.
After that, we quantify the deviations introduced by microlenses to the GW waveforms and show that they would generally not affect events at LIGO/Virgo frequencies.
Finally, we will put our results in an astrophysical context and elaborate on the implications and potential use-cases, especially those related to the localization of GW sources and cosmography.

\section{Methods}

Strongly lensed GWs may be detectable at design sensitivity, based on current lens and binary models~\citep{Ng:2017yiu, Li:2018prc, Oguri:2018muv}.
The typical analysis considers gravitational lensing by a macromodel, such as a galaxy. 
However, microlensing by, e.g., a point-mass within a larger galaxy macromodel may significantly affect the lensed GW waveform \citep{Lai:2018rto, Christian:2018vsi, Jung:2017flg,  Diego:2019lcd, Diego:2019rzc}.
Suppose the wavelength of the GW $\lambda$ is comparable to the Einstein radius of the point mass $\theta_m$. 
In that case, there will be significant diffraction effects, in which case the geometrical optics approximation no longer holds \citep{Takahashi:2003ix}. 
In this section, we review the necessary methods to treat lensing in the full wave-optics regime.
Note: in this paper, we work in units where $c = G = 1$.

\subsection{Computing the Diffraction Integral}\label{sec:diffintegral}
\begin{figure}
        \includegraphics[width=\linewidth]{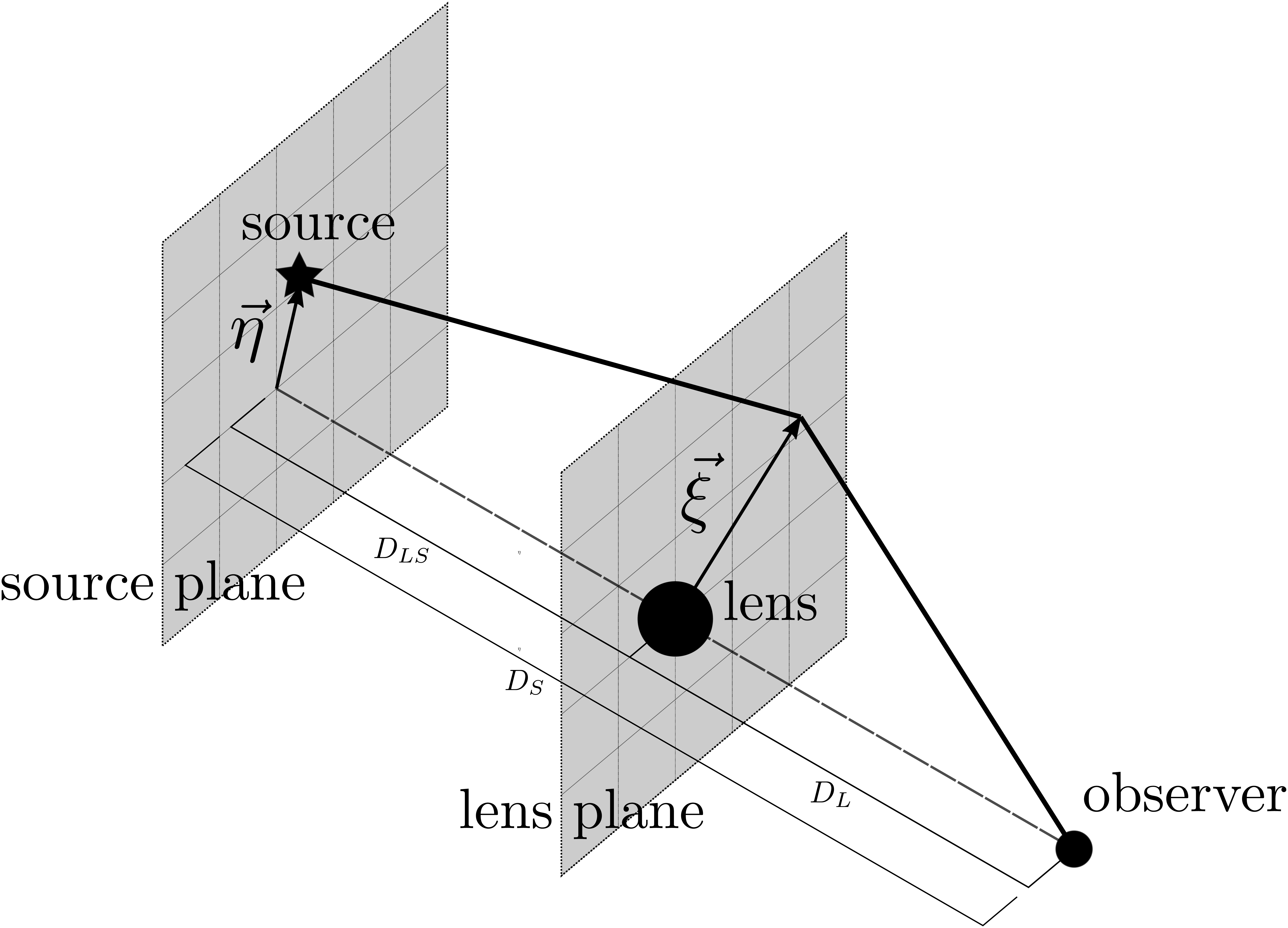}
        \caption{Schematic diagram of the lensing geometry.
        We use the thin-lens approximation and assume that all of the lens mass resides on the lens plane. 
        This diagram is not drawn to scale; the lengths $D_{LS}$, $D_L$ and $D_S$ are very large compared to $\|\vec{\eta}\|$ and $\|\vec{\xi}\|$.
        }
        \label{fig:scheme}
\end{figure}

Figure \ref{fig:scheme} shows the geometry of a typical lensing configuration.
A GW source is positioned at an angular diameter distance of $D_L$. 
In contrast, a massive object, or `lens', is positioned near the straight-line path between the source and the observer at an angular diameter distance $D_S$.
As $D_L$ and $D_S$ are very large compared to the length scale of the lens, we can apply the thin-lens approximation, where the lensing of GWs occurs in the plane of the lens. %

When GWs are radiated from the source, the portion that passes near the lens might be focused toward the observer.
As the lens's length scale is small compared to the distances between the source, lens, and observer, we can approximate the paths of lensed `rays' of GWs as straight line segments that change their direction abruptly at the lens plane.
In Figure \ref{fig:scheme}, we show one of these rays. 
If we draw all possible rays from the source that would arrive at the observer, we would observe a one-to-one correspondence of such rays with points on the lens plane.
Moreover, as these rays take different spacetime paths, they would arrive at the observer at different times.
Thus, we can construct a function $t_d$ that maps a point on the lens plane $\vec{\xi}$ to the arrival time of the ray that passes through that point, given a source position $\vec{\eta}$.
For analytical calculations with $\vec{\xi}$ and $\vec{\eta}$, it is easier to work in units normalized by the Einstein radius
\begin{equation}
\theta_E = \sqrt{\frac{4 M_L D_L D_{LS}}{D_S}},
\end{equation}
where $M_L$ is the total mass of the lens and $D_{LS}$ is the angular diameter distance between the lens and the source (in systems with multiple lenses, we will revert to un-normalized units to avoid confusion).
In these normalized units, writing $\vec{x} = \vec{\xi} / \theta_E$ and $\vec{y} = \vec{\eta} / \theta_E$, the time delay function $t_d$ is given by
\begin{equation}
\label{eq:timedelay}
t_d(\vec{x},\vec{y}) = \frac{D_S \theta_E^2 (1+z_L)}{D_L D_{LS}} T(\vec{x},\vec{y}),
\end{equation}
with
\begin{equation}
T(\vec{x},\vec{y}) = \frac{1}{2}\|\vec{x} - \vec{y}\|^2 - \phi(\vec{x})-T_0(\vec{y}),
\end{equation}
where $T_0(\vec{y})$ corresponds to the arrival time of the first image, and $\phi(\vec{x})$ is the Fermat potential which depends on the nature of the lens (e.g. we have $\phi(\vec{x}) = \ln{\|\vec{x}\|}$ for a point mass lens). 
Besides, we consider the time delay on cosmological distances, such that eq. \ref{eq:timedelay} includes a factor of $1+z_L$, where $z_L$ is the lens's redshift.

Let the GW waveform observed without the presence of the lens be $h(f)$ in the frequency domain.
If the waveform observed with lensing is $h_L(f)$, we can quantify the change in the waveform due to lensing effects by the amplification factor 
\begin{equation}
    F(f) = \dfrac{h_L(f)}{h(f)},
\end{equation}
which is complex in general.
If $\phi$ and hence $t_d$ is known, we can obtain $F(f)$ by the diffraction integral \citep{schneider1992gravitational}
\begin{equation} \label{diffractionintegral}
F(f) = \frac{D_S \theta_E^2 (1+z_L)}{D_L D_{LS}}\frac{f}{i} \iint_{\mathbb{R}^2} d^2\vec{x}\ e^{2 \pi i f t_d(\vec{x},\vec{y})}.
\end{equation}
By using a dimensionless frequency scale $w = 8 \pi M_{L} (1 + z_L) f$, we arrive at
\begin{equation} \label{diffractionintegralnormalized}
\Tilde{F}(w) = -i w \iint_{\mathbb{R}^2} d^2\vec{x}\ e^{i w T(\vec{x},\vec{y})}.
\end{equation}

The integrand in eq. \ref{diffractionintegralnormalized} is highly oscillatory for high $w$, so the integral is computationally expensive to solve with typical numerical integration techniques even for simple lens models.
To solve this integral, we follow a similar method mentioned in \cite{ulmer1994femtolensing} and used in \cite{Diego:2019lcd}, where we compute an area integration directly in the lens plane and take its Fourier transform, which is easier to implement for general lens models and gives $F(f)$ for a wide frequency spectrum with a single calculation.
The case of microlensing on a macromodel introduces important subtleties to this method, making the evaluation of the integral not as straight forward as models containing lenses with masses all of the same order.
In particular, the regions around the time delay function's saddle points require careful treatment (see Sec. \ref{subsec:otherimages}).
These cases and the detailed methodology for calculating eq. \ref{diffractionintegralnormalized} for general lens models will be discussed in a later work (Cheung et al., in preparation).
In this paper, we focus on the results from the minimum time delay image of the macromodel, which can already give us insights into the effects of stellar lensing on GWs.

\subsection{Geometrical optics approximation}

For $w \gg 1$, we can work in the geometrical optics limit, where the integral in eq. \ref{diffractionintegralnormalized} will mostly be contributed by the stationary points of $T(\vec{x},\vec{y})$ (Fermat's principle). 
We call these stationary points the image positions of the lensing configuration.
In such a limit, $\tilde{F}(w)$ is given by \citep{schneider1992gravitational}
\begin{equation} \label{eq:Fgeom}
    \tilde{F}(w) = \sum_j |\mu(\vec{x_j})|^{1/2} \exp{(i \pi (2 w T(\vec{x_j},\vec{y}) - n_j))},
\end{equation}
where $\vec{x_j}$ are the image positions, $\mu(\vec{x}) = \det(\partial \vec{y}/\partial \vec{x})$ is the image magnification, $n_j = 0, \frac{1}{2}, 1$ when $\vec{x_j}$ is at a minimum, saddle, and maximum point of $T(\vec{x},\vec{y})$ respectively.
We solve the image properties in the geometrical optics limit using the \textsc{LensingGW} package~\citep{Pagano:2020rwj}.

\subsection{Microlensing}

\begin{figure}
    \centering
    \includegraphics[width=\linewidth]{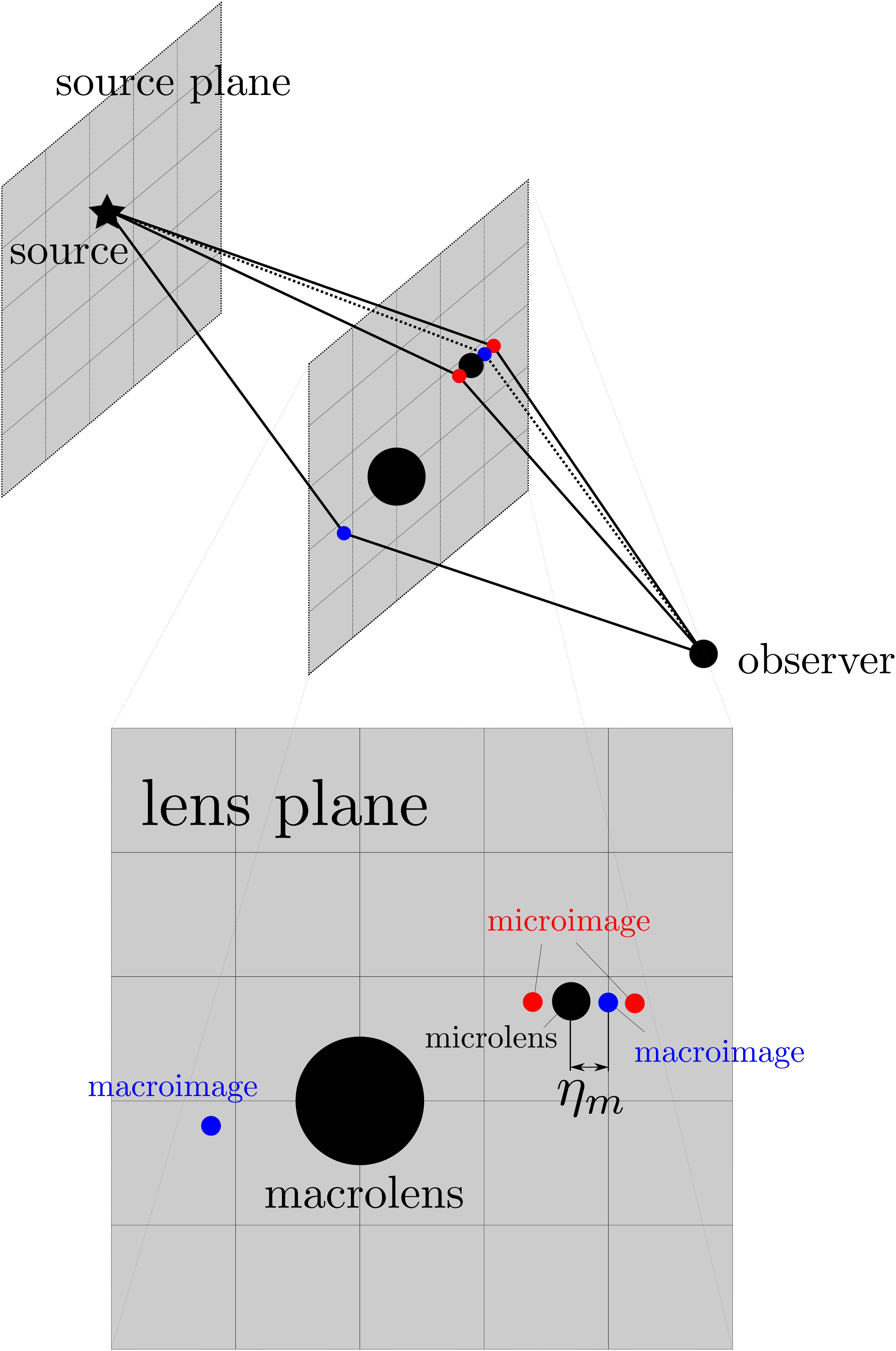}
    \caption{Schematic diagram of an example scenario of microlensing on a macromodel.
    (Top) In the geometrical optics limit, one can treat the waves as rays propagating from the source to the observer, only taking specific routes through particular points on the lens plane.
    If there were no microlens, there would be two of these rays, passing through macroimage points (blue points) on the lens plane (bottom panel) before arriving at the observer.
    However, when a microlens (small black dot on the lens plane) is present near a macroimage, the beam (shown in dotted lines) might be split into multiple ones and pass through different points on the lens plane, which correspond to microimage positions (red points).
    The distance between the microlens and the macroimage is written as $\eta_m$ and called the micro impact parameter.}
    \label{fig:microlensing_geometry}
\end{figure}

Strong lensing occurs when a GW passes by a lens with an impact parameter $\lesssim$ the Einstein radius.
A scenario particularly interesting would be when a GW passes inside the Einstein ring of two lenses of drastically different lens scales. 
It is useful for such cases to introduce the notion of `macro' and `micro' lens models. 
We refer to the lensing images produced by the two models as macroimages and microimages, respectively, following~\cite{Diego:2019lcd}.

Let there be some lenses of mass $\sim M$ and some of mass $\sim m$ in the lens plane (where the thin-lens approximation is invoked), and let $M \gg m$.
For simplicity, let there be no relevant lenses in the intermediate-mass range.
We can then define the $\sim M$ lenses to be the macrolenses, and those $\sim m$ to be the microlenses.
We can obtain an intuitive understanding of the behavior of GWs lensed by such a system by thinking of how the lenses would split the GW beams.
We can think of the macrolenses splitting the beams into separate rays (macroimages), and then the microlenses split the beams further (into microimages). 
An example is shown in Figure \ref{fig:microlensing_geometry}. 

However, there is a caveat: when thinking of lensed GWs as propagating towards us by taking certain discrete ray-like paths, we have implicitly invoked the geometrical optics approximation.
As mentioned previously in this section, if the wavelength of the GWs is comparable to the Einstein radius of the lenses, diffraction effects will occur, and the GWs will no longer take only a discrete number of paths. 
The next section will show that the geometrical optics approximation is insufficient for treating microlenses of stellar mass for GWs detectable by LIGO. 
Nonetheless, the microimage positions and the macroimage positions are always the stationary points of the time delay function $t_d$, and it is convenient to keep track of their positions even if we are working with a full wave-optics treatment.

The Einstein radii of the microlenses are orders of magnitudes smaller than those of the macrolenses. 
The difference in arrival time between a group of microimages near a particular macroimage is also orders of magnitude below the difference in arrival time between macroimages. 
Therefore, we can carry out our integration in Eq.~\ref{diffractionintegralnormalized}  for the microimages near the minimum macroimage of the macromodel (galactic lens), that is, within a small domain on the lens plane near the macroimage. 
We will then obtain an amplification factor $F(f)$ that corresponds to the interferences between the relevant microimages with diffraction included.
For saddle point macroimages, we will have to cover a much more extensive integration domain.
This is to be performed in a later work (see also Sec. \ref{subsec:otherimages}).

\section{Results}

This section quantifies the effects of stellar microlenses on GWs (within the LIGO frequency band) lensed by a galaxy.
We place the source at a redshift of $z_S = 1$ and place a galactic lens at $z_L = 0.5$.
For our galactic macromodel, we use a singular isothermal sphere (SIS) lens of mass $10^{10} \, \rm M_{\odot}$.
We will only consider cases corresponding to strong lensing. The macro impact parameter $y$ (position of the source in the sky with respect to the center of the SIS macromodel) is less than $\theta_\mathit{SIS}$, the Einstein radius of the SIS macromodel.
For such cases, we will see two macroimages, one corresponding to a minimum point in the time delay function, and the other corresponding to the saddle point.
Then, we introduce point-mass microlenses in the vicinity of the minimum macroimage.
We can naively treat the minimum macroimage position as the `source position' for the microlensing configuration and define the micro impact parameter $\eta_m$ as the distance of the microlens from the macroimage position.

We assume that the microlens is hosted by the macromodel galaxy, which will likely be true for most cases of microlensing where the microlens is a star from the nuclear cluster of the galaxy, so we can treat the macrolens and the microlens as residing on the same plane in the sky \citep{Diego:2019lcd}. 
If the microlenses are of stellar mass, say $m < 100 M_{\odot}$, the arrival time of the microimages will reside within a time window which is less than the period of GWs measured by LIGO. Thus, we will have to use the methods discussed in the previous section to recover the amplification factor for the signal measured that corresponds to the lensed GW rays near the macroimage.

In this section, as we are working with two very different length scales corresponding to the macromodel and microlenses, we will revert to unnormalized units of length, i.e., using $\vec{\xi}$ and $\vec{\eta}$ instead of $\vec{x}$ and $\vec{y}$.
We denote the norms of the vectors $\xi = \|\vec{\xi}\|$ and $\eta = \|\vec{\eta}\|$ etc.
Relevant lengths for the SIS macromodel are the macro impact parameter (distance from the center of the SIS lens to the source, projected onto the lens plane) $\eta_\mathit{SIS}$, and the Einstein radius of the SIS lens $\theta_\mathit{SIS}$.
For microlenses near a parent macroimage, we denote the micro impact parameter (distance between the parent macroimage and the microlens on the lens plane) as $\eta_m$, and the Einstein radius of the microlens as $\theta_m$.

\subsection{The necessity of a full wave-optics analysis}

\begin{figure*}
        \includegraphics[width=\linewidth]{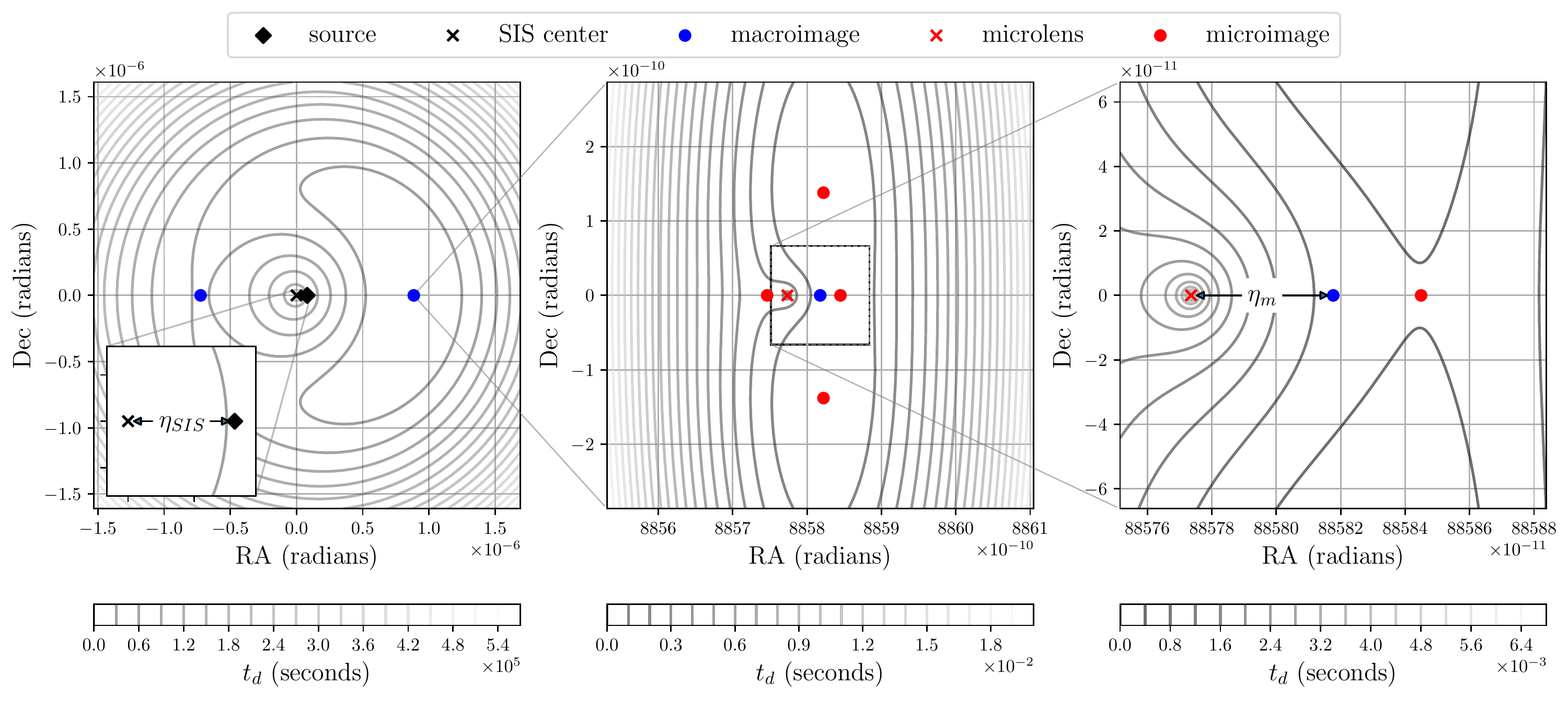}
        \caption{The lens plane of an SIS macromodel of mass $M = 10^{10} \, \rm M_{\odot}$ with a microlens of mass $m = 30 \, \rm M_{\odot}$ near the minimum macroimage. 
        The source position measured from the center of the SIS lens and projected onto the lens plane (macro impact parameter) is $\eta_\mathit{SIS} = 0.1 \, \theta_\mathit{SIS}$, while the microlens is placed $\eta_m = 1.0 \, \theta_m$ away from the macroimage.
        The contour lines of the time delay function $t_d$ are shown in grey.
        (Left panel) view of the complete model, including the source position and macroimages.
        The macroimage on the left is of type-II (negative parity) as it is a saddle point of $t_d$, while that on the right is of type-I (positive parity).
        (Center panel) zoom in on the region around the type-I minimum macroimage. 
        The microlens splits the macroimage into four microimages, two of type-I (both minimum points) and two of type-II.
        (Right panel) Further zoom in onto the macroimage position.
        Note that as the macroimage is split into microimages, the point corresponding to the macroimage position is not a stationary point of $t_d$ anymore; it is an actual image only when the microlens is absent.}
        \label{fig:a_contour}
\end{figure*}

Figure \ref{fig:a_contour} shows the configuration of the full lens model and the contours of the time delay function for the case of a macro impact parameter $\eta_\mathit{SIS} = 0.1 \, \theta_\mathit{SIS}$, micro impact parameter $\eta_m = 1.0 \, \theta_{m}$, and microlens mass $m = 30 \, \rm M_{\odot}$. 
The microlens is placed on the straight line between the SIS center and the minimum macroimage position.
More general cases are studied in Appendix \ref{appen:double} and \ref{sec:appa}.
The macroimage and microimage positions are found with the help of the \textsc{LensingGW} package \citep{Pagano:2020rwj}.

We can work in the geometrical optics limit for the macromodel (the SIS lens) due to the lens's considerable mass. 
We will observe two GW signals of the same phase but different amplitudes and arrival times (see the two macroimages in the left panel of Figure \ref{fig:a_contour}). 
In this case, the macroimage only induces an overall magnification on the waveform, such that $F(f)$ is a constant. 

\begin{figure}
        \includegraphics[width=\linewidth]{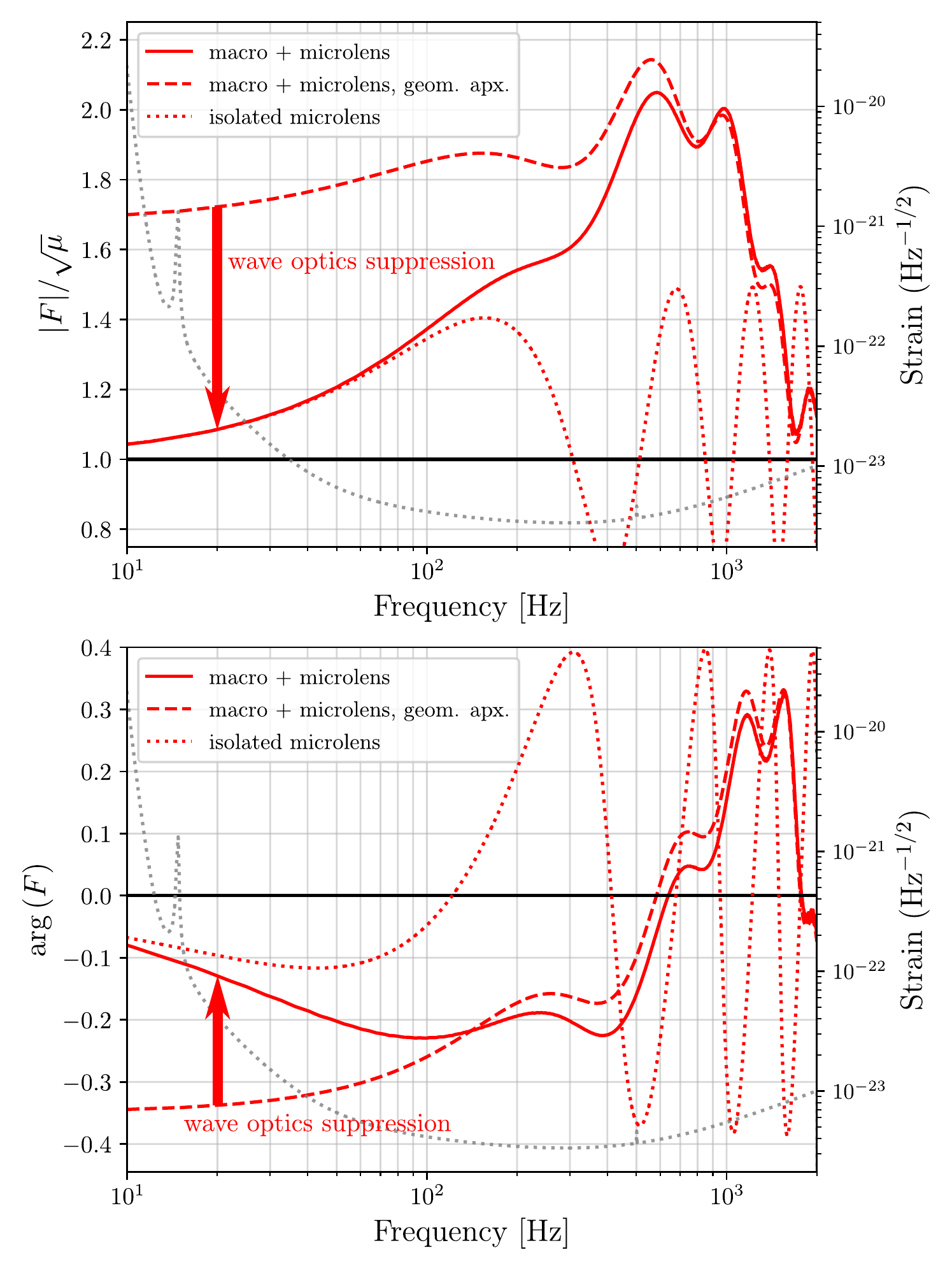}
        \caption{The normalized norm (top, left axis) and phase (bottom, left axis) of the frequency domain amplification factor $F(f)$ near the minimum macroimage for the lensing set up described in Figure \ref{fig:a_contour} ($30 \, \rm M_\odot$ microlens).
        The deviation of the geometrical optics approximation (red dashed line) and the full wave optics treatment (red solid line) demonstrates that wave optics effects are important for stellar lenses of lower masses, and they suppress the amplification factor.
        The disagreement between treating the microlens in isolation (red dotted line) and considering both the macro and microlenses (red solid line) shows that we have to also consider the macromodel's contribution to the time delay function in our analysis.
        The PSD of LIGO is plotted in grey dotted lines with the axis on the right of each panel.}
        \label{fig:a_30F}
\end{figure}

However, the addition of a stellar-mass lens near the macroimage disturbs the time delay function near the arrival time of the macroimage, causing $F(f)$ to deviate from the constant value. 
Figure \ref{fig:a_30F} shows $F(f)$ for the GW rays that come from around the minimum macroimage. 
Note that the amplification factor is normalized by $\sqrt{\mu}$, where $\mu$ is the magnification of the macroimage in the geometrical optics limit.
If we work with the geometrical optics approximation and use eq. \ref{eq:Fgeom}, we obtain the dashed curve, which deviates significantly from the results without approximations for lower frequencies, showing that a full wave-optics analysis is necessary when dealing with microlenses.

We also demonstrate that if we naively treated the microlenses independently of the macromodel (in isolation), we obtain the dotted curve in Figure~\ref{fig:a_30F} instead, corresponding to the standard analytical expression for the amplification factor of the point mass lens (see~\cite{peters1974index}). 
This amplification factor is significantly different from the full non-approximate amplification factor. 
In fact, by observing that the microlens splits the macroimage into four microimages instead of two in the center panel of Figure \ref{fig:a_contour}, we can already infer that the macromodel would introduce qualitative differences from the case of an isolated point mass.
This further confirms the results in~\cite{Diego:2019lcd,Diego:2019rzc,Pagano:2020rwj}, that it is necessary to consider the macromodel when treating microlensing in the presence of strong lensing. 

\subsection{Waveform suppression due to diffraction}

\begin{figure}
        \centering
        \includegraphics[width=\linewidth]{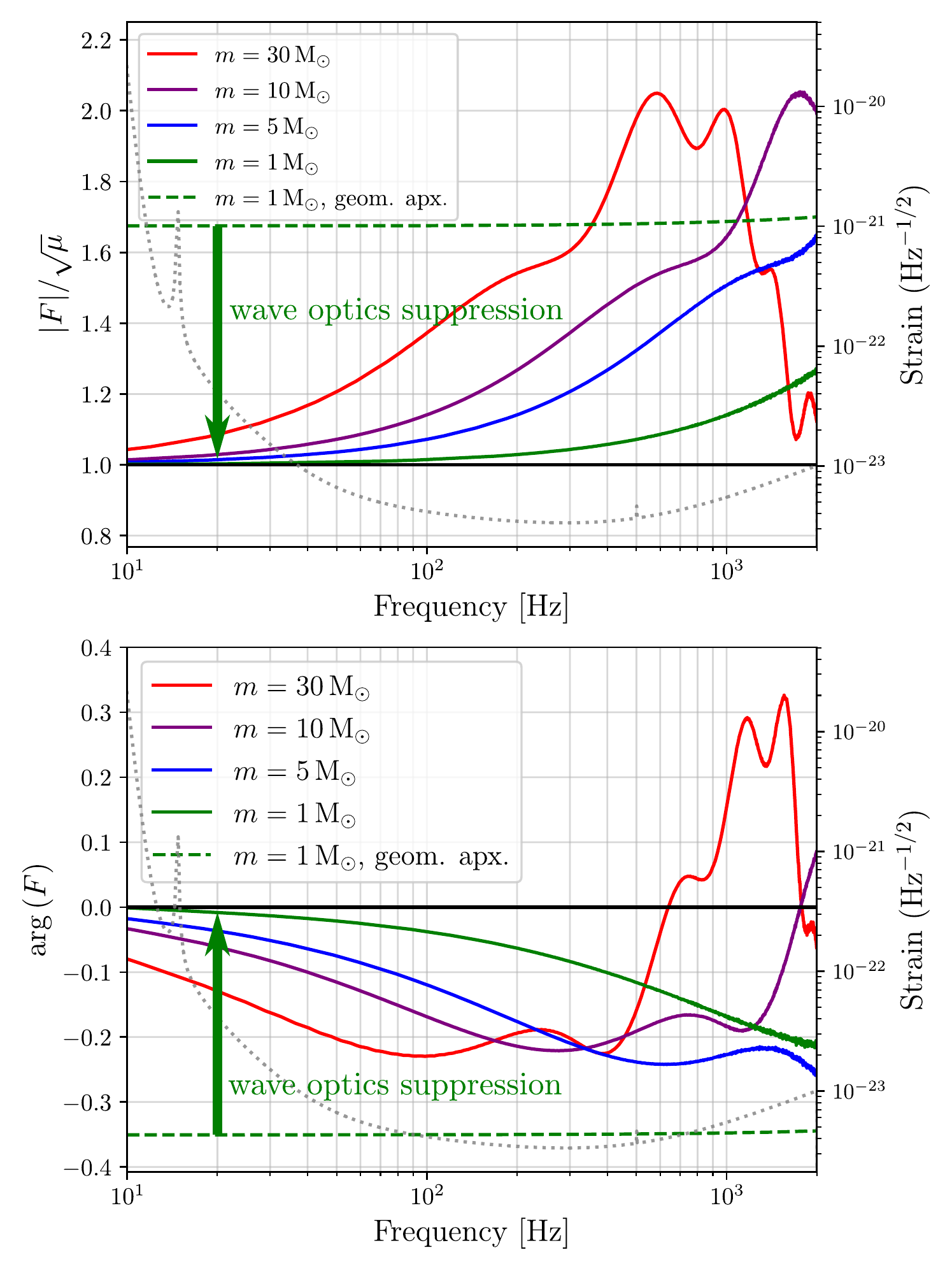}
        \caption{Same as Figure \ref{fig:a_30F}, but with a point mass microlens of mass $m \in \{1, 5, 10, 30\} \, \rm M_{\odot}$, placed accordingly at a distance of $\eta_m = 1.0 \, \theta_m$ away from the macroimage position similar to the set up in Figure \ref{fig:a_contour}.
        $F/\sqrt{\mu}$ converges to the constant limit of $1$ for low frequencies and low microlens masses. 
        We show the geometrical optics approximation results for a $1 \, \rm M_\odot$ microlens in the green dashed lines. 
        The wave optics diffraction effects suppress the waveform relative to the geometrical optics approximation. 
        }
        \label{fig:a2}
\end{figure}

\begin{figure}
        \centering
        \includegraphics[width=\linewidth]{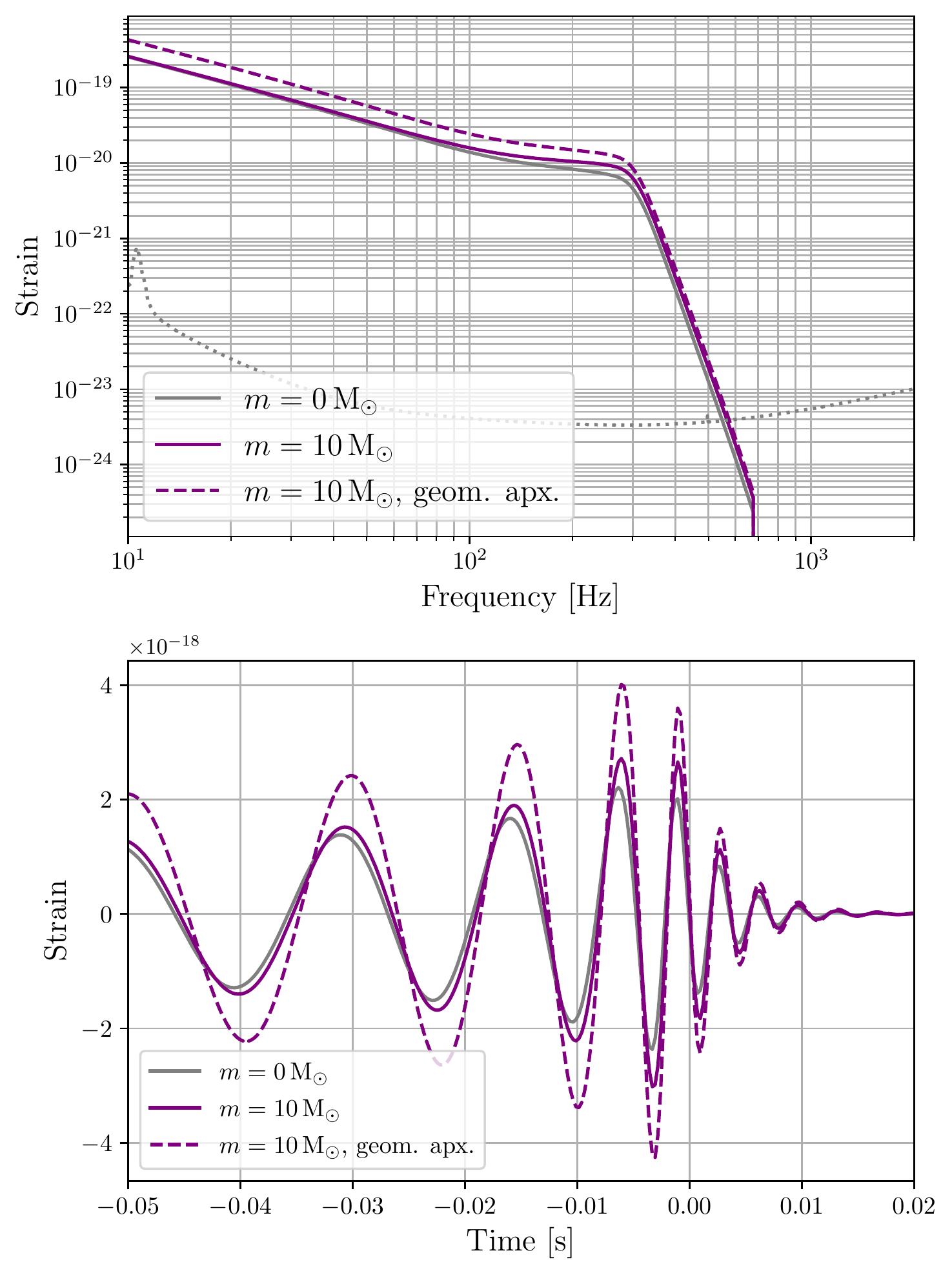}
        \caption{The frequency domain (top) and time domain (bottom) waveforms for the cases $m \in \{0, 10\} \rm M_\odot$ for the set up shown in Figure \ref{fig:a_contour}. 
        The waveform obtained for $m = 10 \, \rm M_\odot$ from the geometrical optics limit is plotted in dashed lines.
        $m = 0 \, \rm M_\odot$ corresponds to the case where there is no microlens.
        The PSD of aLIGO is also plotted in the top panel in black dotted lines for reference.}
        \label{fig:a_waveform}
\end{figure}

We saw that the normalized amplification factor in our full wave-optics analysis reduces to unity as $f \rightarrow 0$, which corresponds to the limit when the wavelength of the GW $\lambda > \theta_{m}$, where the waves would propagate through the microlens unimpeded as diffraction effects dominate.
This effectively means that GWs will be largely unaffected by the microlenses in the very low-mass regime.
We examine the level of waveform suppression for different microlens masses in Figure \ref{fig:a2}.
With $m \in \{1, 5, 10, 30\} \, \rm M_{\odot}$, we see that lower microlens masses correspond to amplification factors approaching $1$ at most relevant LIGO/Virgo frequencies, implying that the GW signal is amplified only by the macroimage magnification.
We also plot the frequency domain and time domain waveforms for $m \in \{0, 10\} \, \rm M_{\odot}$ in Figure \ref{fig:a_waveform}.
Here, an IMRPhenomD waveform of a $30 + 30 \, \rm M_{\odot}$ binary black hole merger generated by \textsc{pycbc} \citep{pycbc} is used.
The result obtained from the geometrical optics approximation for the $m = 10 \, \rm M_\odot$ waveform is also shown in dashed lines, showing that we would overestimate the deviations of the waveform due to microlensing if we work in the geometrical optics limit.

\begin{figure}
        \centering
        \includegraphics[width=\linewidth]{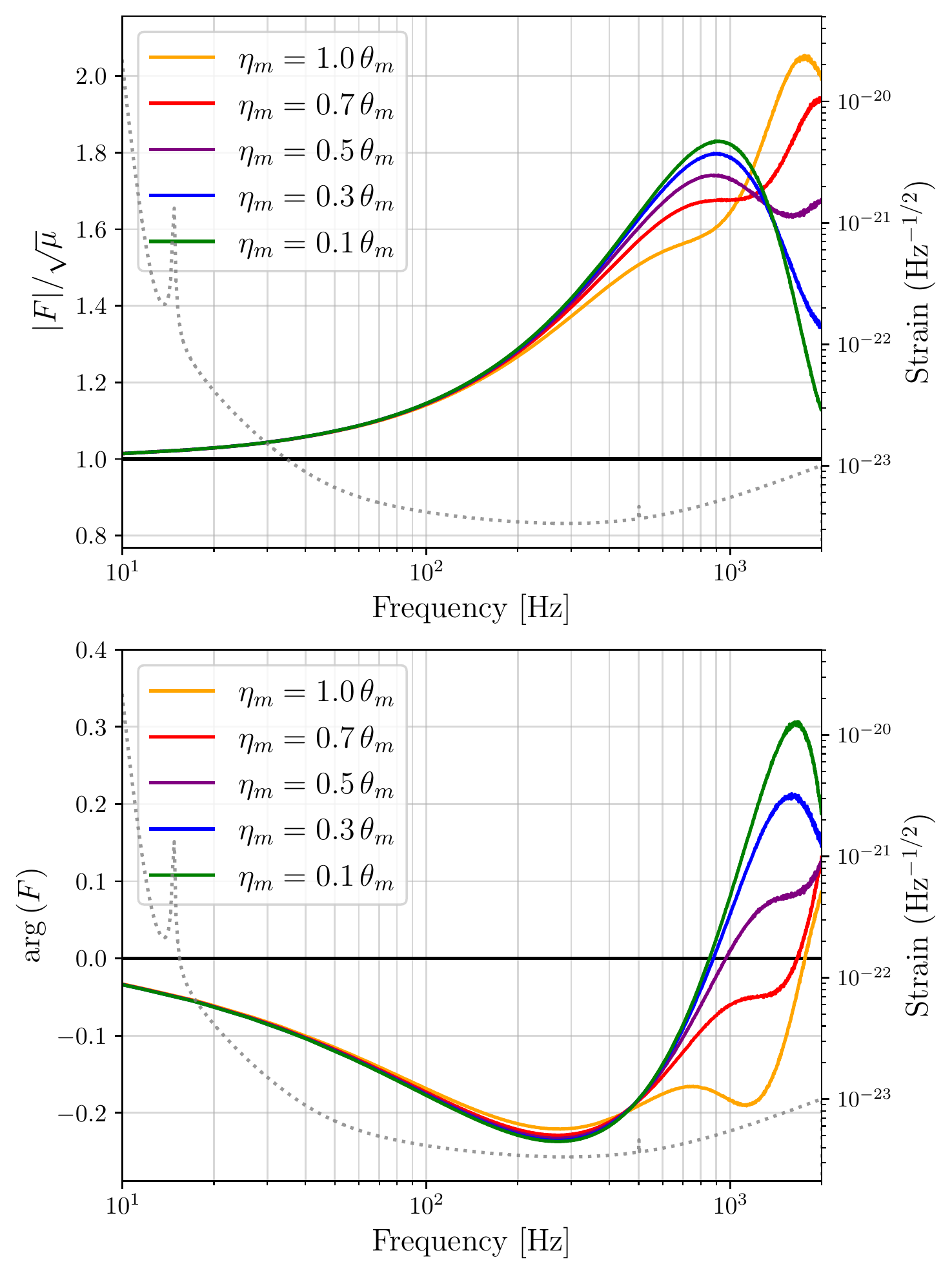}
        \caption{Same as Figure \ref{fig:a2}, but this time fixing the microlens mass at $m = 10 \, \rm M_\odot$ while varying the micro impact parameter $\eta_m \in \{0.1, 0.3, 0.5, 0.7, 1.0\} \, \theta_{m}$.
        Due to diffraction effects, the amplification factors in the lower frequency region are suppressed to the same limit regardless of the micro impact parameter.}
        \label{fig:e}
\end{figure}

Then, we investigate how the microlens location relative to the macroimage affects waveform suppression.
Fixing $m = 10 \, \rm M_{\odot}$, we vary the micro impact parameter $\eta_m \in \{0.1, 0.3, 0.5, 0.7, 1.0\} \, \theta_{m}$ and calculate the amplification factors at different image positions.
As shown in Figure \ref{fig:e}, the amplification factor converges to the same curve at low frequencies, suggesting that the effects of waveform suppression are similar regardless of the microlens location.
This behavior is also observed in the case of isolated point mass lenses \citep{Takahashi:2003ix, Nakamura:1997sw}. 

\begin{figure}
    \centering
    \includegraphics[width = \linewidth]{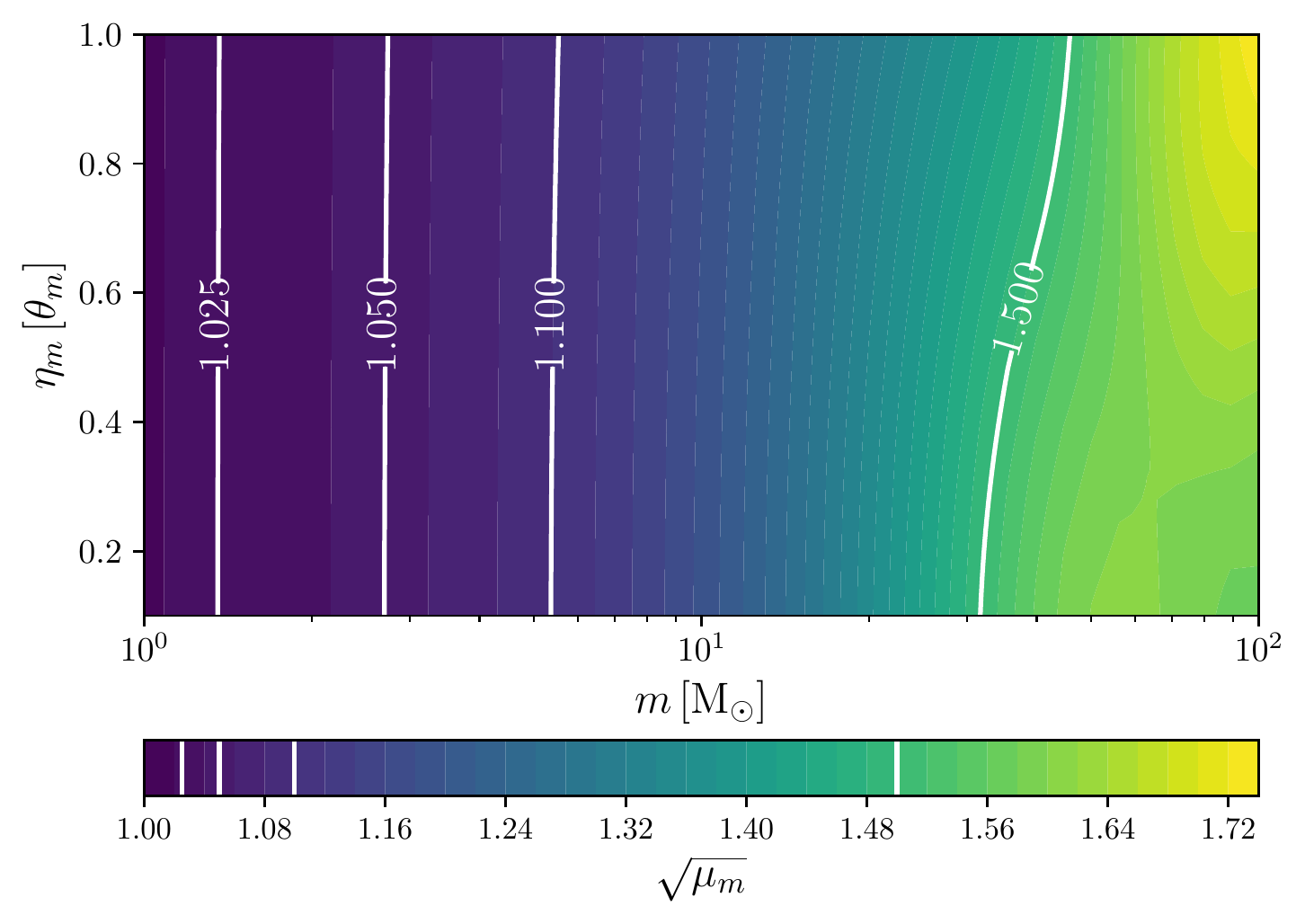}
    \caption{Contour plot of the effective micro-amplificaiton $\sqrt{\mu_m}$ for our lens set up as a function of the microlens mass $m$ and the micro impact parameter $\eta_m$.
    The behavior of $\sqrt{\mu_m}$ almost does not depend on $\eta_m$ for a low $m$.
    }
    \label{fig:micromu}
\end{figure}

We then quantify the differences introduced by microlensing to the lensed waveforms.
We first define 
\begin{align}
    &h_{L,\mathit{full}}(f) = F(f)h(f),\\
    &h_{L, \mathit{macro}}(f) = \sqrt{\mu} \, h(f).
\end{align}
Then, we define the effective micro-magnification by
\begin{equation} \label{eq:effmu}
    \mu_m = \frac{(h_{L,\mathit{full}}\,|\,h_{L,\mathit{full}})}{(h_{L,\mathit{macro}}\,|\,h_{L,\mathit{macro}})},
\end{equation}
where 
\begin{equation}
    (a|b) = 4 \operatorname{Re} \int^{f_\mathit{high}}_{f_\mathit{low}} \frac{a(f)b^*(f)}{S_n(f)}
\end{equation}
is the inner product weighted by the power-spectral density (PSD) $S_n(f)$ of LIGO.
By taking the noise weighted inner product, we remove the frequency dependence, so we can now approximate the change induced by the microlens on the amplification of the GW waveform.
In Figure \ref{fig:micromu}, we plot $\sqrt{\mu_m}$ for $m \in (1, 100) \, \rm M_\odot$ and $\eta_m \in (0.1, 1.0) \, \theta_m$.
For $h(f)$, we again use an IMRPhenomD waveform of a $30 + 30 \, \rm M_{\odot}$ binary black hole merger, generated with \textsc{pycbc} \citep{pycbc}.
We took the square root because the amplification to the waveform is $\sim \sqrt{\mu_m}$.
If the microlens were absent, $h_{L,\mathit{full}} = h_{L, \mathit{macro}}$ and eq. \ref{eq:effmu} reduces to $\mu_m = 1$.
For lower $m$, $\mu_m$ does not significantly deviate from unity.
For $m \lesssim 3 \, \rm M_\odot$, $\sqrt{\mu_m}$ is larger than $1$ by $\lesssim 5\%$.

To further quantify the deviation in the waveforms, we define the waveform overlap, which is given by
\begin{equation}
    \mathcal{O}[h_{L,\mathit{macro}}, h_{L,\mathit{full}}] = \frac{(h_{L,\mathit{full}}\,|\,h_{L,\mathit{macro}})}{\sqrt{(h_{L,\mathit{full}}\,|\,h_{L,\mathit{full}})(h_{L,\mathit{macro}}\,|\,h_{L,\mathit{macro}})}},
\end{equation}
We then define the match $\mathcal{M}[h_{L,\mathit{macro}}, h_{L,\mathit{full}}]$ to be the maximum of $\mathcal{O}[h_{L,\mathit{macro}}, h_{L,\mathit{full}}]$ over time and phase, and $1 - \mathcal{M}[h_{L,\mathit{macro}}, h_{L,\mathit{full}}]$ to be the microlens induced mismatch.
The match is also computed using the \textsc{pycbc} software \citep{pycbc}.

\begin{figure}
        \centering
        \includegraphics[width=\linewidth]{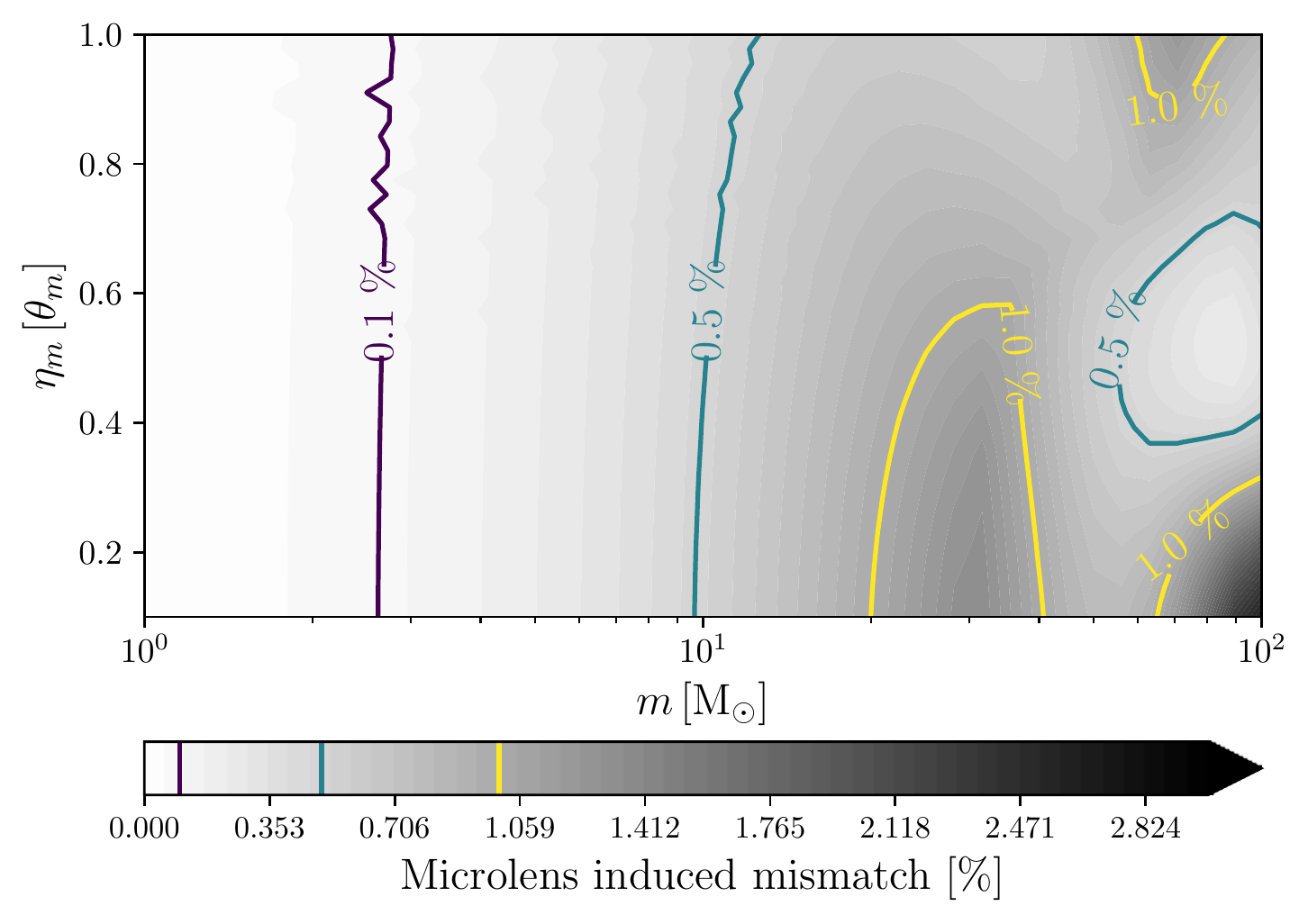}
        \caption{The mismatch between the waveforms with and without microlensing for our lens set up as a function of the microlens mass $m$ and the micro impact parameter $\eta_m$.
        The contour lines of mismatch $\in \{0.1, 0.5, 1.0\} \% $ are shown.
        Stellar mass microlenses of $m \lesssim 10 \, \rm M_\odot$ have a mismatch in the subpercent level.}
        \label{fig:match}
\end{figure}

Figure \ref{fig:match} shows a contour plot of the mismatch $1 - \mathcal{M}[h_{L,\mathit{macro}}, h_{L,\mathit{full}}]$ between the waveforms with and without microlensing as a function of $m$ and $\eta_m$. 
For most of the parameter space concerned, the mismatch $\lesssim 1 \%$ except for $m, \eta_m \sim 100 \, \rm M_\odot, 0.1 \, \theta_m$.
For $m \lesssim 10 \, \rm M_\odot$, the mismatch $\lesssim 0.5 \%$.
We note that the waveform's mismatch does not encode any information about the macroimage magnification $\mu$, which scales the signal to noise ratio (SNR) of the signal as $\sqrt{\mu}$.
However, as the mismatch is at the sub-percent level for typical stellar masses of $m \lesssim 10 \, \rm M_\odot$, it is expected that microlenses would not significantly affect the measurement of other GW parameters for typical magnifications $\mu \lesssim 10$. 
We note that more complicated scenarios with a field of microlenses still exist and there are some rare cases when we could observe extremely large macromodel magnifications ($\mu \sim 100$). 
These scenarios may still induce more substantial stellar microlensing features on the waveform~\citep{Diego:2019lcd,Diego:2019rzc}. 

\section{Discussion} \label{sec:discussion}

\subsection{Implications of waveform suppression}

In most micro- and millilensing situations above $100 \, \rm M_\odot$, the geometrical optics limit describes the physical configuration well at the most sensitive LIGO/Virgo frequencies.
However, for microlenses below $100 \, \rm M_\odot$, wave diffraction effects become non-negligible.
Wave optics effects suppress the lensing magnification due to microlenses, especially in common astrophysical strong lensing scenarios with microlenses $\sim 1 \, \rm M_\odot$ (Figure~\ref{fig:a2}). 
We note that there are still rarer astrophysical situations where two or more microlenses could plausibly conspire together to mimic larger lenses~\citep{Diego:2019lcd,Diego:2019rzc}, which is discussed in Appendix \ref{appen:double}.
However, our results are nevertheless promising, 
and may have exciting implications for studies of fundamental physics and cosmology through localization studies. 

In particular, as pointed out by \cite{oguri2019strong}, typical microlensing may significantly and adversely affect the accuracy and precision of time delay and magnification measurements of lensing transients. 
As the typical radii of the objects emitting gravitational waves are small (the orbital radius of binary compact object mergers can be as low as $R \sim 10 \, \rm km$), GWs can in principle be subject to considerable microlensing in the usual geometrical optics limit~\citep{oguri2019strong}. 
However, our results confirm that the wave optics effects suppress stellar microlensing (see also discussion on isolated lenses in~\cite{oguri2019strong}).  
This suggests that gravitational waves may allow for relatively clean measurements where we need to worry less about microlensing systematics in, e.g., localization studies~\citep{hannuksela2020localizing} and possible follow-up studies thereafter~\citep{Sereno:2011ty,Collett:2016dey,Baker:2016reh,Fan:2016swi,Liao:2017ioi,Cao:2019kgn,Li:2019rns,hannuksela2020localizing}.
Indeed, for example, the lensed GWs may be a promising means for us to constrain cosmological parameters, such as the Hubble constant $H_0$, by accurately measuring the time delay of different strongly lensed GW images. 
Intriguingly, GWs are not as vulnerable to microlensing as the EM waves from supernovae and other explosive transients (see discussion in~\cite{oguri2019strong}). 
Systematic errors due to microlensing can potentially play a significant role in these measurements. 
Thus, wave optics suppression can be seen as an advantage in this case.

Moreover, \cite{hannuksela2020localizing} demonstrated the use of quadruply lensed GWs to localize the host galaxy by matching the GW image properties with the lensed galaxy systems found in the direction of the GW in an electromagnetic follow-up. 
Microlensing could induce a significant error on the lensing time-delay and magnification, hampering the match between the image properties (magnifications, time delays) found in the EM and the GW channels. 
Our work suggests that these localization studies may suffer from stellar microlensing in the common astrophysical scenarios to a much lesser degree, which may spell improvements for these localization studies. 
Subsequently, microlensing may pose less of a problem in measuring $H_0$ through methods based on the distance-ladder or the time-delay distance. 
However, there is still more work in investigating more complex microlensing scenarios, such as fields of microlenses near the saddle point macroimage (type-II image).
Nevertheless, our results are very promising. 

Finally, our results illustrate the mass ranges for which wave optics effect becomes significant. 
As shown in Figure \ref{fig:a2}, both the magnitude and argument of the amplification factor deviate significantly from the geometric optics model even at a microlens mass of $30 \, \rm M_\odot$.
In Appendix \ref{appen:geom}, we show that the validity of the geometrical optics approximation depends at least on the mass of the lens and the frequency range in question, and the approximation might be good enough for LIGO detections for microlenses of mass $m \gtrsim 300 \, \rm M_\odot$.

\subsection{Non-minimum macroimages} \label{subsec:otherimages}
In this paper, we focused on the amplification factors of microlenses near the macromodel minimum point (a type-I image) of the time delay function.
This is for computational simplicity, as only the area immediately surrounding the minimum contributes to the diffraction integral around the macroimage time delay.

However, for a type-II images (saddle points), the regions of the lens plane with time delay $t_d(\vec{\xi},\vec{\eta}) \in [t_{\mathit{im}} - \tau, t_{\mathit{im}} + \tau]$ correspond to thin strips centering around the contour lines $t_d(\vec{\xi},\vec{\eta}) = t_{\mathit{im}}$.
As the macroimage is a saddle point, such a contour line can have an arbitrary length that depends on the lens model.
In SIS lensing with typical galactic masses, the relevant contours at the saddle point will be vast, making the calculation of eq. \ref{diffractionintegralnormalized} very computationally expensive.
It has been shown in \cite{Diego:2019lcd} that microlensing around type-II images have qualitative differences from that around type-I ones. 
We will develop methods to tackle type-II images of general lens configurations in future work.

\section{Conclusions}
We have shown that a wave optics treatment of the full lens model is required for stellar-mass microlensing on a galactic macromodel. 
Due to diffraction effects, microlensing is suppressed, making the contributions by stellar lenses of mass $\sim 1 \, \rm M_\odot$ small regardless of their proximity to the macrolensed GW beams.
This allows for more accurate measurements of the magnification and time delay of lensed GW images and better confidence in localization studies. 
On the flip side, we might require third-generation detectors to use GW lensing to probe objects of masses $\mathcal{O}(1 \, \rm M_\odot)$. %
Future work will include the cases of microlensing near type-II macroimages.

\section*{Acknowledgements}

We would like to thank Apratim Ganguly, Ajit Mehta, Anupreeta More, Eungwang Seo, and K. Haris for their insightful comments. 
Furthermore, we would like to thank Miguel Zumalacarregui for helping with the calculation of the diffraction integral, and also Thomas Collett for related discussions. 
OAH is supported by the research program of the Netherlands Organization for Scientific Research (NWO).
TGFL is partially supported by grants from the Research Grants Council of Hong Kong (Project No. 14306218), Research Committee of the Chinese University of Hong Kong, and the Croucher Foundation of Hong Kong.
We are grateful for computational resources provided by the LIGO Laboratory and supported by the National Science Foundation Grants PHY-0757058 and PHY-0823459.  

\section*{Data Availability}

The data underlying this article will be shared on reasonable request to the corresponding authors.

\bibliographystyle{mnras}
\bibliography{wolensing} %

\begin{thebibliography}{}
\makeatletter
\relax
\def\mn@urlcharsother{\let\do\@makeother \do\$\do\&\do\#\do\^\do\_\do\%\do\~}
\def\mn@doi{\begingroup\mn@urlcharsother \@ifnextchar [ {\mn@doi@}
  {\mn@doi@[]}}
\def\mn@doi@[#1]#2{\def\@tempa{#1}\ifx\@tempa\@empty \href
  {http://dx.doi.org/#2} {doi:#2}\else \href {http://dx.doi.org/#2} {#1}\fi
  \endgroup}
\def\mn@eprint#1#2{\mn@eprint@#1:#2::\@nil}
\def\mn@eprint@arXiv#1{\href {http://arxiv.org/abs/#1} {{\tt arXiv:#1}}}
\def\mn@eprint@dblp#1{\href {http://dblp.uni-trier.de/rec/bibtex/#1.xml}
  {dblp:#1}}
\def\mn@eprint@#1:#2:#3:#4\@nil{\def\@tempa {#1}\def\@tempb {#2}\def\@tempc
  {#3}\ifx \@tempc \@empty \let \@tempc \@tempb \let \@tempb \@tempa \fi \ifx
  \@tempb \@empty \def\@tempb {arXiv}\fi \@ifundefined
  {mn@eprint@\@tempb}{\@tempb:\@tempc}{\expandafter \expandafter \csname
  mn@eprint@\@tempb\endcsname \expandafter{\@tempc}}}

\bibitem[\protect\citeauthoryear{Baker \& Trodden}{Baker \&
  Trodden}{2017}]{Baker:2016reh}
Baker T.,  Trodden M.,  2017, \mn@doi [Phys. Rev. D]
  {10.1103/PhysRevD.95.063512}, 95, 063512

\bibitem[\protect\citeauthoryear{Cao, Qi, Cao, Biesiada, Li, Pan  \& Zhu}{Cao
  et~al.}{2019}]{Cao:2019kgn}
Cao S.,  Qi J.,  Cao Z.,  Biesiada M.,  Li J.,  Pan Y.,   Zhu Z.-H.,  2019,
  \mn@doi [Sci. Rep.] {10.1038/s41598-019-47616-4}, 9, 11608

\bibitem[\protect\citeauthoryear{Chen \& Holz}{Chen \&
  Holz}{2016}]{Chen:2016tys}
Chen H.-Y.,  Holz D.~E.,  2016, arXiv preprint arXiv:1612.01471

\bibitem[\protect\citeauthoryear{Christian, Vitale  \& Loeb}{Christian
  et~al.}{2018}]{Christian:2018vsi}
Christian P.,  Vitale S.,   Loeb A.,  2018, \mn@doi [Phys. Rev. D]
  {10.1103/PhysRevD.98.103022}, 98, 103022

\bibitem[\protect\citeauthoryear{Collett \& Bacon}{Collett \&
  Bacon}{2017}]{Collett:2016dey}
Collett T.~E.,  Bacon D.,  2017, \mn@doi [Phys. Rev. Lett.]
  {10.1103/PhysRevLett.118.091101}, 118, 091101

\bibitem[\protect\citeauthoryear{Dai \& Venumadhav}{Dai \&
  Venumadhav}{2017}]{Dai:2017huk}
Dai L.,  Venumadhav T.,  2017, arXiv preprint arXiv:1702.04724

\bibitem[\protect\citeauthoryear{Dai, Zackay, Venumadhav, Roulet  \&
  Zaldarriaga}{Dai et~al.}{2020}]{Dai:2020tpj}
Dai L.,  Zackay B.,  Venumadhav T.,  Roulet J.,   Zaldarriaga M.,  2020, arXiv
  preprint arXiv:2007.12709

\bibitem[\protect\citeauthoryear{Deguchi \& Watson}{Deguchi \&
  Watson}{1986}]{Deguchi:1986zz}
Deguchi S.,  Watson W.~D.,  1986, \mn@doi [Phys. Rev. D]
  {10.1103/PhysRevD.34.1708}, 34, 1708

\bibitem[\protect\citeauthoryear{Diego}{Diego}{2020}]{Diego:2019rzc}
Diego J.~M.,  2020, \mn@doi [Phys. Rev. D] {10.1103/PhysRevD.101.123512}, 101,
  123512

\bibitem[\protect\citeauthoryear{Diego, Hannuksela, Kelly, Broadhurst, Kim, Li,
  Smoot  \& Pagano}{Diego et~al.}{2019}]{Diego:2019lcd}
Diego J.,  Hannuksela O.,  Kelly P.,  Broadhurst T.,  Kim K.,  Li T.,  Smoot
  G.,   Pagano G.,  2019, \mn@doi [Astron. Astrophys.]
  {10.1051/0004-6361/201935490}, 627, A130

\bibitem[\protect\citeauthoryear{Ezquiaga, Holz, Hu, Lagos  \& Wald}{Ezquiaga
  et~al.}{2020}]{Ezquiaga:2020gdt}
Ezquiaga J.~M.,  Holz D.~E.,  Hu W.,  Lagos M.,   Wald R.~M.,  2020, arXiv
  preprint arXiv:2008.12814

\bibitem[\protect\citeauthoryear{Fan, Liao, Biesiada, Piorkowska-Kurpas  \&
  Zhu}{Fan et~al.}{2017}]{Fan:2016swi}
Fan X.-L.,  Liao K.,  Biesiada M.,  Piorkowska-Kurpas A.,   Zhu Z.-H.,  2017,
  \mn@doi [Phys. Rev. Lett.] {10.1103/PhysRevLett.118.091102}, 118, 091102

\bibitem[\protect\citeauthoryear{Goyal, Haris, Mehta  \& Ajith}{Goyal
  et~al.}{2020}]{Goyal:2020bkm}
Goyal S.,  Haris K.,  Mehta A.~K.,   Ajith P.,  2020, arXiv preprint
  arXiv:2008.07060

\bibitem[\protect\citeauthoryear{Hannuksela, Haris, Ng, Kumar, Mehta, Keitel,
  Li  \& Ajith}{Hannuksela et~al.}{2019}]{Hannuksela:2019kle}
Hannuksela O.,  Haris K.,  Ng K.,  Kumar S.,  Mehta A.,  Keitel D.,  Li T.,
  Ajith P.,  2019, \mn@doi [Astrophys. J. Lett.] {10.3847/2041-8213/ab0c0f},
  874, L2

\bibitem[\protect\citeauthoryear{Hannuksela, Collett, Çalışkan  \&
  Li}{Hannuksela et~al.}{2020}]{hannuksela2020localizing}
Hannuksela O.~A.,  Collett T.~E.,  Çalışkan M.,   Li T. G.~F.,  2020,
  \mn@doi [Monthly Notices of the Royal Astronomical Society]
  {10.1093/mnras/staa2577}, 498, 3395

\bibitem[\protect\citeauthoryear{Haris, Mehta, Kumar, Venumadhav  \&
  Ajith}{Haris et~al.}{2018}]{Haris:2018vmn}
Haris K.,  Mehta A.~K.,  Kumar S.,  Venumadhav T.,   Ajith P.,  2018, arXiv
  preprint arXiv:1807.07062

\bibitem[\protect\citeauthoryear{Jung \& Shin}{Jung \&
  Shin}{2019}]{Jung:2017flg}
Jung S.,  Shin C.~S.,  2019, \mn@doi [Phys. Rev. Lett.]
  {10.1103/PhysRevLett.122.041103}, 122, 041103

\bibitem[\protect\citeauthoryear{Lai, Hannuksela, Herrera-Mart{\'\i}n, Diego,
  Broadhurst  \& Li}{Lai et~al.}{2018}]{Lai:2018rto}
Lai K.-H.,  Hannuksela O.~A.,  Herrera-Mart{\'\i}n A.,  Diego J.~M.,
  Broadhurst T.,   Li T.~G.,  2018, \mn@doi [Phys. Rev. D]
  {10.1103/PhysRevD.98.083005}, 98, 083005

\bibitem[\protect\citeauthoryear{Li, Mao, Zhao  \& Lu}{Li
  et~al.}{2018}]{Li:2018prc}
Li S.-S.,  Mao S.,  Zhao Y.,   Lu Y.,  2018, \mn@doi [Mon. Not. Roy. Astron.
  Soc.] {10.1093/mnras/sty411}, 476, 2220

\bibitem[\protect\citeauthoryear{Li, Lo, Sachdev, Chan, Lin, Li  \&
  Weinstein}{Li et~al.}{2019a}]{Li:2019osa}
Li A.~K.,  Lo R.~K.,  Sachdev S.,  Chan C.,  Lin E.,  Li T.~G.,   Weinstein
  A.~J.,  2019a, arXiv preprint arXiv:1904.06020

\bibitem[\protect\citeauthoryear{Li, Fan  \& Gou}{Li
  et~al.}{2019b}]{Li:2019rns}
Li Y.,  Fan X.,   Gou L.,  2019b, \mn@doi [Astrophys. J.]
  {10.3847/1538-4357/ab037e}, 873, 37

\bibitem[\protect\citeauthoryear{Liao, Fan, Ding, Biesiada  \& Zhu}{Liao
  et~al.}{2017}]{Liao:2017ioi}
Liao K.,  Fan X.-L.,  Ding X.-H.,  Biesiada M.,   Zhu Z.-H.,  2017, \mn@doi
  [Nature Commun.] {10.1038/s41467-017-01152-9}, 8, 1148

\bibitem[\protect\citeauthoryear{Liu, Hernandez  \& Creighton}{Liu
  et~al.}{2020}]{Liu:2020par}
Liu X.,  Hernandez I.~M.,   Creighton J.,  2020, arXiv preprint
  arXiv:2009.06539

\bibitem[\protect\citeauthoryear{McIsaac, Keitel, Collett, Harry, Mozzon, Edy
  \& Bacon}{McIsaac et~al.}{2020}]{McIsaac:2019use}
McIsaac C.,  Keitel D.,  Collett T.,  Harry I.,  Mozzon S.,  Edy O.,   Bacon
  D.,  2020, \mn@doi [Phys. Rev. D] {10.1103/PhysRevD.102.084031}, 102, 084031

\bibitem[\protect\citeauthoryear{Nakamura}{Nakamura}{1998}]{Nakamura:1997sw}
Nakamura T.~T.,  1998, \mn@doi [Phys. Rev. Lett.]
  {10.1103/PhysRevLett.80.1138}, 80, 1138

\bibitem[\protect\citeauthoryear{Ng, Wong, Broadhurst  \& Li}{Ng
  et~al.}{2018}]{Ng:2017yiu}
Ng K.~K.,  Wong K.~W.,  Broadhurst T.,   Li T.~G.,  2018, \mn@doi [Phys. Rev.
  D] {10.1103/PhysRevD.97.023012}, 97, 023012

\bibitem[\protect\citeauthoryear{Nitz et~al.,}{Nitz et~al.}{2020}]{pycbc}
Nitz A.,  et~al., 2020, gwastro/pycbc: PyCBC release v1.16.11,
  \mn@doi{10.5281/zenodo.4134752}, \url
  {https://doi.org/10.5281/zenodo.4134752}

\bibitem[\protect\citeauthoryear{Oguri}{Oguri}{2018}]{Oguri:2018muv}
Oguri M.,  2018, \mn@doi [Mon. Not. Roy. Astron. Soc.] {10.1093/mnras/sty2145},
  480, 3842

\bibitem[\protect\citeauthoryear{Oguri}{Oguri}{2019}]{oguri2019strong}
Oguri M.,  2019, \mn@doi [Reports on Progress in Physics]
  {10.1088/1361-6633/ab4fc5}, 82, 126901

\bibitem[\protect\citeauthoryear{Ohanian}{Ohanian}{1974}]{Ohanian:1974ys}
Ohanian H.,  1974, \mn@doi [Int. J. Theor. Phys.] {10.1007/BF01810927}, 9, 425

\bibitem[\protect\citeauthoryear{Pagano, Hannuksela  \& Li}{Pagano
  et~al.}{2020}]{Pagano:2020rwj}
Pagano G.,  Hannuksela O.~A.,   Li T.~G.,  2020, \mn@doi [Astron. Astrophys.]
  {10.1051/0004-6361/202038730}, 643, A167

\bibitem[\protect\citeauthoryear{Pang, Hannuksela, Dietrich, Pagano  \&
  Harry}{Pang et~al.}{2020}]{Pang:2020qow}
Pang P.~T.,  Hannuksela O.~A.,  Dietrich T.,  Pagano G.,   Harry I.~W.,  2020,
  \mn@doi [Mon. Not. Roy. Astron. Soc.] {10.1093/mnras/staa1430}

\bibitem[\protect\citeauthoryear{Peters}{Peters}{1974}]{peters1974index}
Peters P.~C.,  1974, \mn@doi [Phys. Rev. D] {10.1103/PhysRevD.9.2207}, 9, 2207

\bibitem[\protect\citeauthoryear{Robertson, Smith, Massey, Eke, Jauzac,
  Bianconi  \& Ryczanowski}{Robertson et~al.}{2020}]{Robertson:2020mfh}
Robertson A.,  Smith G.~P.,  Massey R.,  Eke V.,  Jauzac M.,  Bianconi M.,
  Ryczanowski D.,  2020, \mn@doi [Mon. Not. Roy. Astron. Soc.]
  {10.1093/mnras/staa1429}

\bibitem[\protect\citeauthoryear{Ryczanowski, Smith, Bianconi, Massey,
  Robertson  \& Jauzac}{Ryczanowski et~al.}{2020}]{Ryczanowski:2020mlt}
Ryczanowski D.,  Smith G.~P.,  Bianconi M.,  Massey R.,  Robertson A.,   Jauzac
  M.,  2020, \mn@doi [Mon. Not. Roy. Astron. Soc.] {10.1093/mnras/staa1274},
  495, 1666

\bibitem[\protect\citeauthoryear{Schneider, Ehlers  \& Falco}{Schneider
  et~al.}{1992}]{schneider1992gravitational}
Schneider P.,  Ehlers J.,   Falco E.~E.,  1992, Gravitational Lenses.
Springer

\bibitem[\protect\citeauthoryear{Sereno, Jetzer, Sesana  \& Volonteri}{Sereno
  et~al.}{2011}]{Sereno:2011ty}
Sereno M.,  Jetzer P.,  Sesana A.,   Volonteri M.,  2011, \mn@doi [Mon. Not.
  Roy. Astron. Soc.] {10.1111/j.1365-2966.2011.18895.x}, 415, 2773

\bibitem[\protect\citeauthoryear{Smith et~al.}{Smith
  et~al.}{2017}]{Smith:2018gle}
Smith G.,  et~al., 2017, \mn@doi [IAU Symp.] {10.1017/S1743921318003757}, 338,
  98

\bibitem[\protect\citeauthoryear{Smith, Jauzac, Veitch, Farr, Massey  \&
  Richard}{Smith et~al.}{2018}]{Smith:2017mqu}
Smith G.~P.,  Jauzac M.,  Veitch J.,  Farr W.~M.,  Massey R.,   Richard J.,
  2018, \mn@doi [Mon. Not. Roy. Astron. Soc.] {10.1093/mnras/sty031}, 475, 3823

\bibitem[\protect\citeauthoryear{Smith, Robertson, Bianconi  \& Jauzac}{Smith
  et~al.}{2019}]{Smith:2019dis}
Smith G.~P.,  Robertson A.,  Bianconi M.,   Jauzac M.,  2019, arXiv preprint
  arXiv:1902.05140

\bibitem[\protect\citeauthoryear{Takahashi \& Nakamura}{Takahashi \&
  Nakamura}{2003}]{Takahashi:2003ix}
Takahashi R.,  Nakamura T.,  2003, \mn@doi [Astrophys. J.] {10.1086/377430},
  595, 1039

\bibitem[\protect\citeauthoryear{Thorne}{Thorne}{1982}]{Thorne:1982cv}
Thorne K.,  1982, in {Les Houches Summer School on Gravitational Radiation}. pp
  1--57

\bibitem[\protect\citeauthoryear{Ulmer \& Goodman}{Ulmer \&
  Goodman}{1994}]{ulmer1994femtolensing}
Ulmer A.,  Goodman J.,  1994, arXiv preprint astro-ph/9406042

\bibitem[\protect\citeauthoryear{Wang, Stebbins  \& Turner}{Wang
  et~al.}{1996}]{Wang:1996as}
Wang Y.,  Stebbins A.,   Turner E.~L.,  1996, \mn@doi [Phys. Rev. Lett.]
  {10.1103/PhysRevLett.77.2875}, 77, 2875

\bibitem[\protect\citeauthoryear{Yu, Zhang  \& Wang}{Yu
  et~al.}{2020}]{Yu:2020agu}
Yu H.,  Zhang P.,   Wang F.-Y.,  2020, \mn@doi [Mon. Not. Roy. Astron. Soc.]
  {10.1093/mnras/staa1952}, 497, 204

\makeatother
\end{thebibliography}

\appendix

\section{Accuracy of the geometrical optics approximation}
\label{appen:geom}

\begin{figure}
    \centering
    \includegraphics[width=\linewidth]{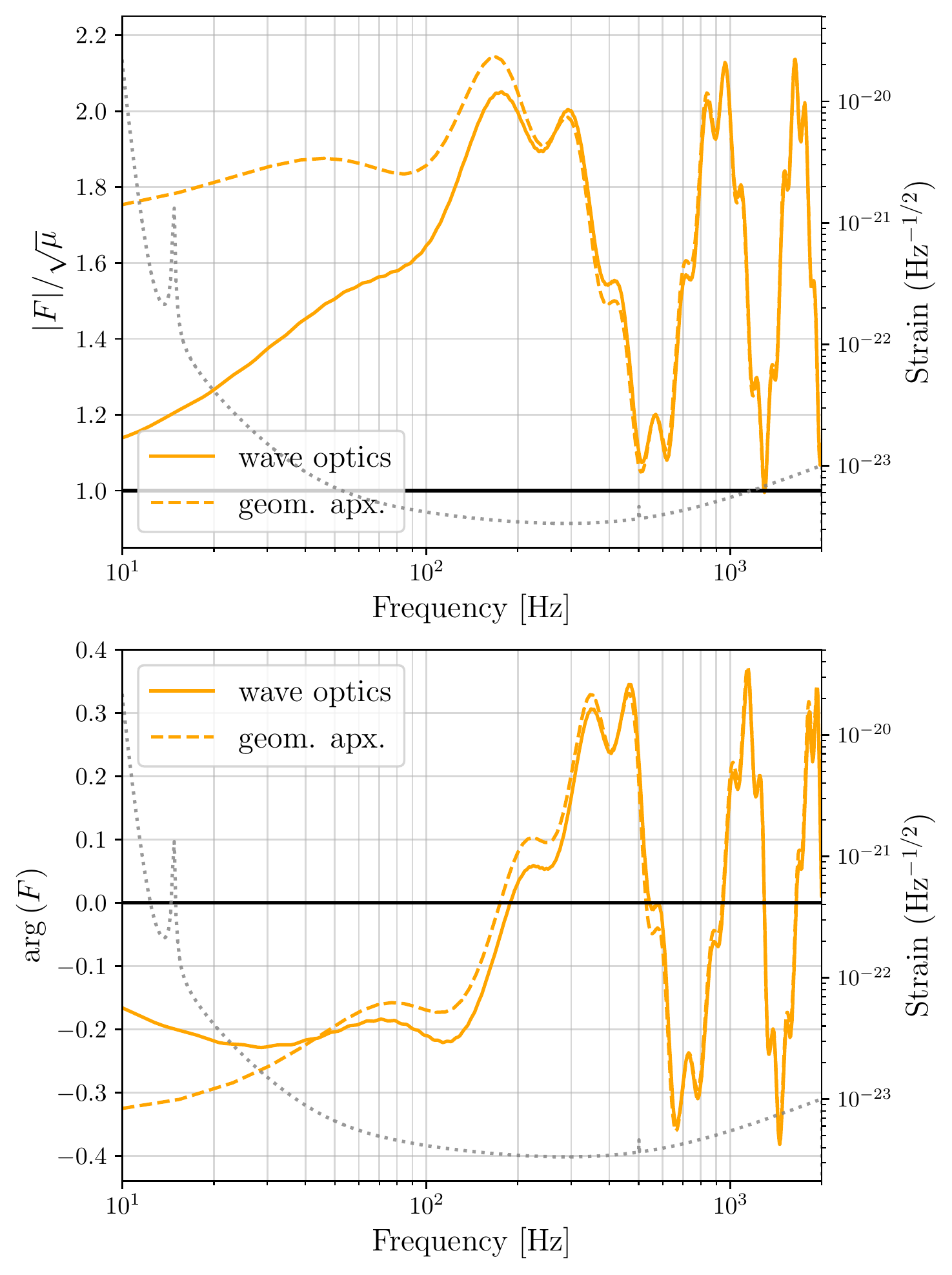}
    \caption{Same as Figure \ref{fig:a_30F}, but with a microlens mass of $m = 100 \, \rm M_\odot$ instead.
    There is still a clear mismatch between the geometrical and wave optics limits for $f \lesssim 10^2 \, \rm Hz$.}
    \label{fig:Fs_100}
\end{figure}

\begin{figure}
    \centering
    \includegraphics[width=\linewidth]{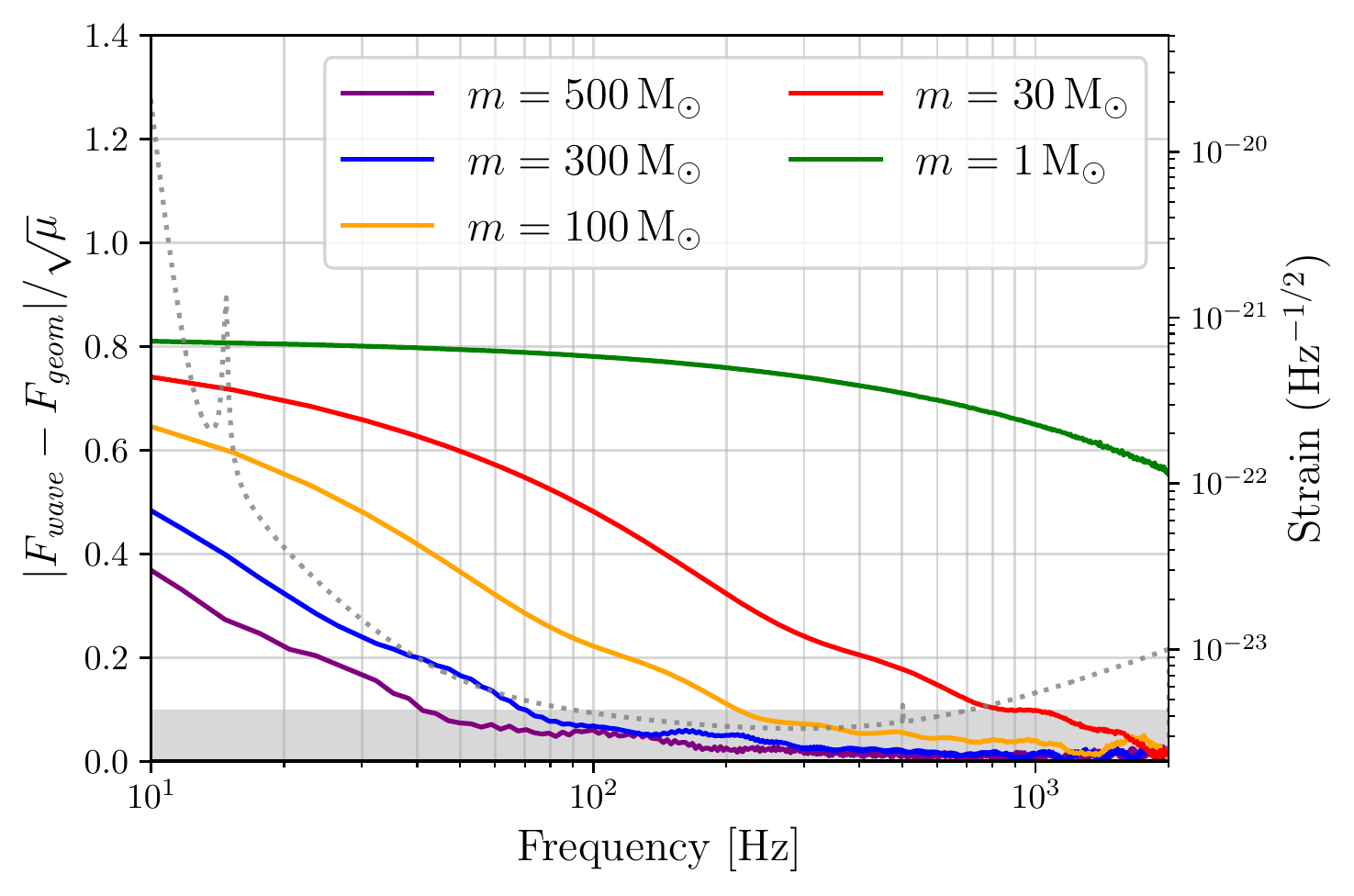}
    \caption{Deviation between the wave optics and geometrical optics results of the amplification factor for $m \in \{1, 30, 100, 300, 500\} \, \rm M_\odot$.
    The shaded region corresponds to $|F_\mathit{wave}-F_\mathit{geom}|/\sqrt{\mu} < 0.1$.
    The curves for $m \in \{300, 500\} \, \rm M_\odot$ lie wtihin this region for the most sensitive frequency range $f \sim \mathcal{O}(10^2 \, \rm Hz)$ of LIGO.}
    \label{fig:dev}
\end{figure}

It has generally been accepted that the geometrical optics limit approximates the exact results better for higher frequencies $f$ and higher (micro)lens masses $m$. 
In Figure \ref{fig:Fs_100}, we compare the full wave optics calculation (solid lines) and geometrical optics approximate (dashed lines) of the amplification factor $F$ for the set up in Figure \ref{fig:a_contour} but with $m = 100 \, \rm M_\odot$. 
Comparing Figures \ref{fig:a_30F} and \ref{fig:Fs_100}, we see that the geometrical optics approximation approaches the wave optics calculations at a lower frequency for the more massive case $m = 100 \, \rm M_\odot$.
More explicitly, we can quantify the error of the geometrical optics limit by taking the norm of the difference between the wave optics amplification factor $F_\mathit{wave}$ and the geometrical optics amplification factor $F_\mathit{geom}$.
In Figure \ref{fig:dev} we plot the results for $m \in \{1, 30, 100, 300, 500\} \, \rm M_\odot$ (normalized by $\sqrt{\mu}$).
For $m \in \{300, 500\} \, \rm M_\odot$, the normalized error in the geometrical optics approximation is $< 0.1$ for most of the frequency range sensitive for LIGO, meaning that the approximation might be good enough for $m \gtrsim 300 \, \rm M_\odot$ depending on the accuracy required.

\section{Two microlenses near the macroimage}
\label{appen:double}

\begin{figure}
    \centering
    \includegraphics[width = .9\linewidth]{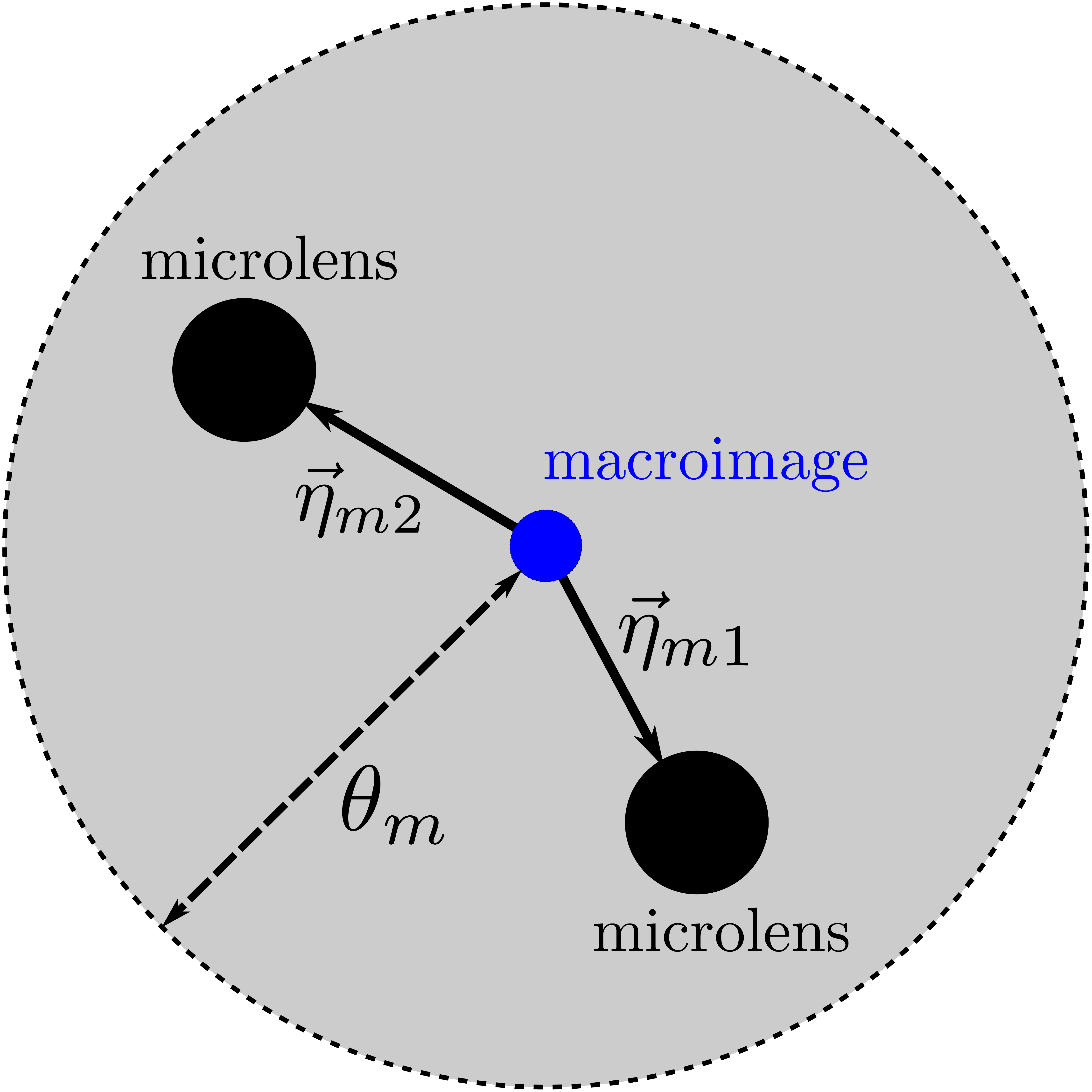}
    \caption{Microlensing set up for the two microlenses case. 
    We place two microlenses, both of mass $m = 1 \, \rm M_\odot$, at random positions $\vec{\eta}_{m1}$ and $\vec{\eta}_{m2}$ within an Einstein radius $\theta_m$ from the macroimage.}
    \label{fig:double_scheme}
\end{figure}

\begin{figure}
    \centering
    \includegraphics[width=\linewidth]{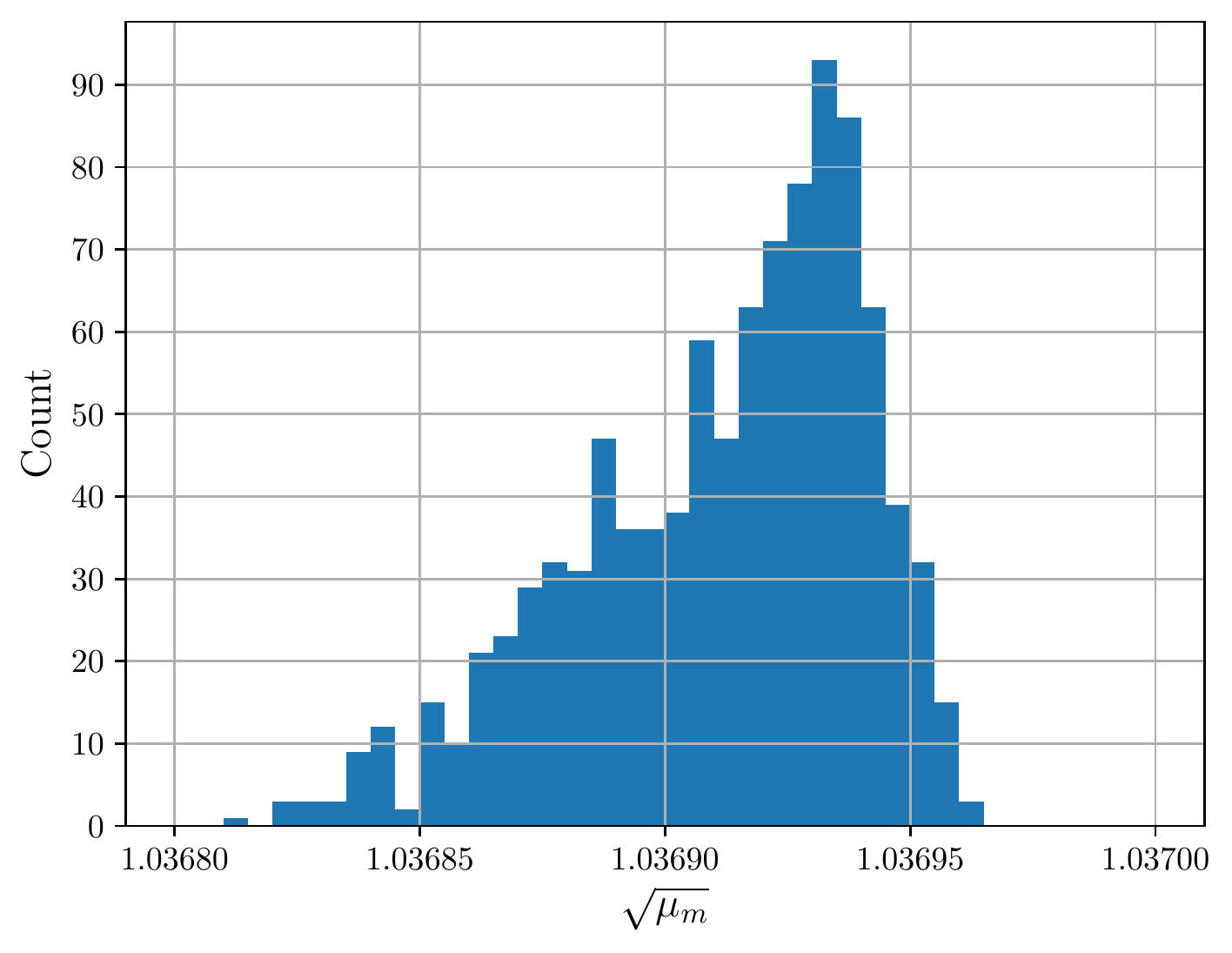}
    \includegraphics[width=\linewidth]{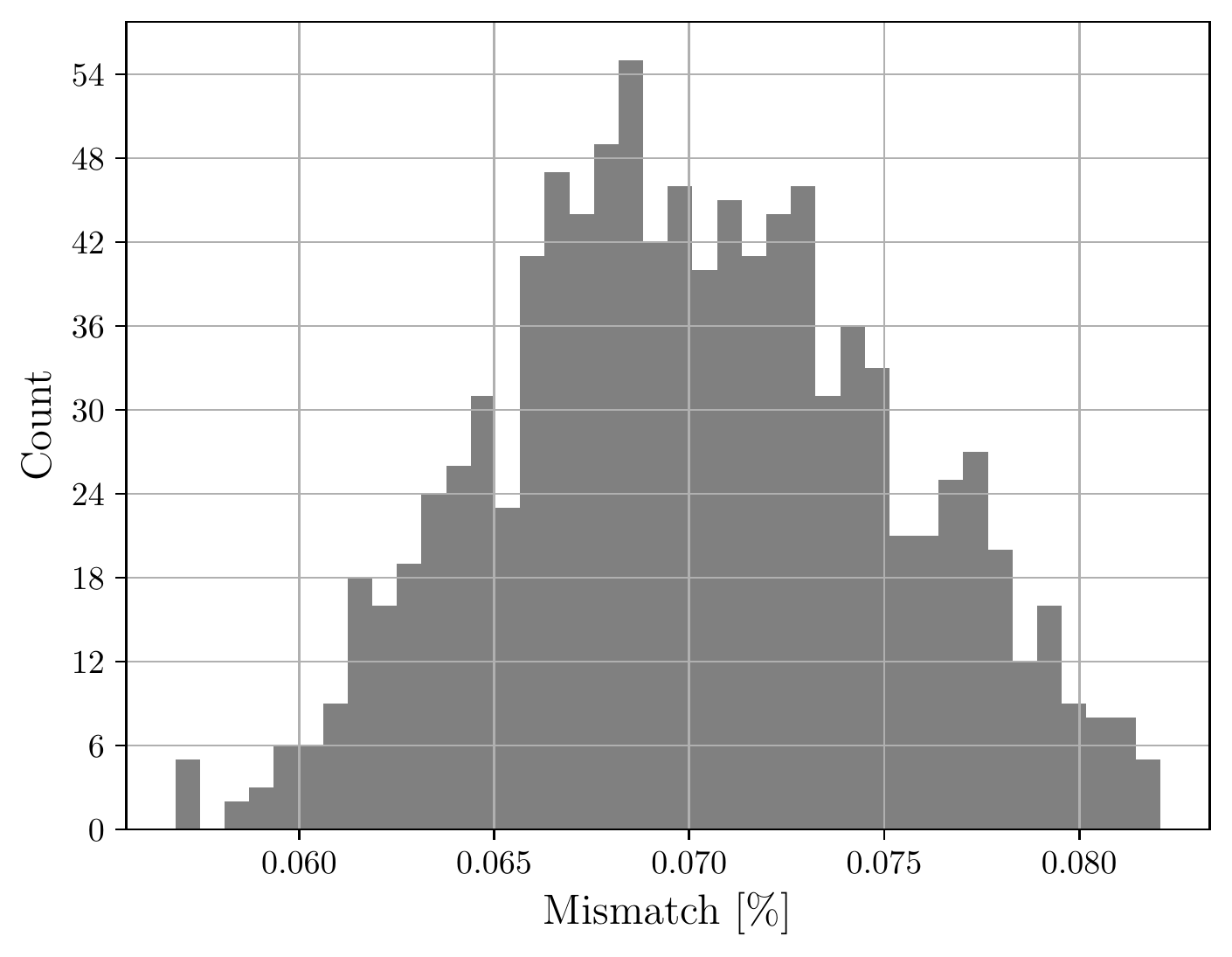}
    \caption{The effective micro amplification $\sqrt{\mu_m}$ and microlens induced mismatch for the $1000$ simulated double microlens configurations.
    $\sqrt{\mu_m} < 1.04$ and mismatch $< 0.1 \%$ for all of these configurations.}
    \label{fig:double}
\end{figure}

Although microlensing due to a single star might not affect GWs significantly, we are interested in the case of GW microlensing by multiple stars as well.
Here we study the case of a GW ray passing through the Einstein radii of two stars.
We use the same $10^{10} \, \rm M_\odot$ SIS macromodel as that in the main text, and restrict ourselves to the minimum macroimage.
As shown in Figure \ref{fig:double_scheme}, we place two microlenses, both of mass $1 \, \rm M_\odot$, at random positions $\vec{\eta}_{m1}$ and $\vec{\eta}_{m2}$ within an Einstein radius $\theta_m$ of the macroimage ($\theta_m$ corresponds to the Einstein radius of a lens with mass $m = 1 \, \rm M_\odot$).
We simulate $1000$ of these microlens pairs, calculate the amplification factor for each configuration, and compute the micro magnification factor as in eq. \ref{eq:effmu} and the microlens induced mismatch as in Figure \ref{fig:match}.
The results are shown in Figure \ref{fig:double}.
The effective micro amplification $\sqrt{\mu_m}$ increases when compared to the single microlens case (see Figure \ref{fig:micromu}), but it still deviates little from unity, with $\sqrt{\mu_m} < 1.04$.
Likewise, the waveform mismatch increases only slightly due to the additional microlens, remaining in the $< 0.1 \%$ range.
This shows that even for microlensing by two stars, the effects are still relatively small.
In general, however, we find that both the microlensing magnification and the induced mismatch increase with additional microlenses. 

\section{Microlenses placed at other angles from the macroimage}
\label{sec:appa}

\begin{figure}
    \centering
    \includegraphics[width=.9\linewidth]{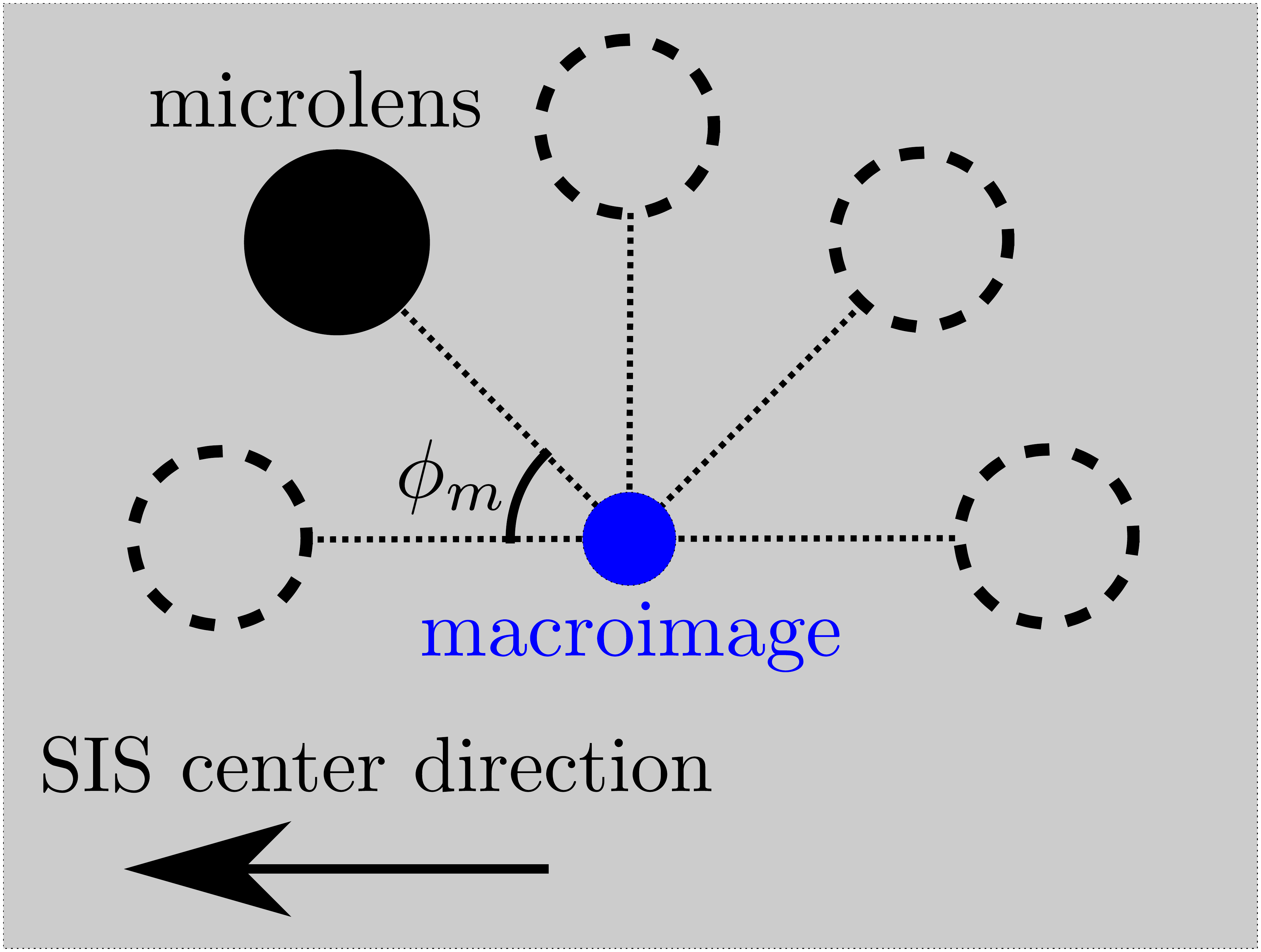}
    \caption{Varying the angular direction with respect to the macroimage for which we place the microlens. 
    The SIS center is to the left of the figure and $\phi_m$ is the angle between the microlens and the SIS center measured at the macroimage.
    The dashed and solid circles correspond to $\phi_m \in \{0\degree, 45\degree, 90\degree, 135\degree, 180\degree\}$.}
    \label{fig:angle_scheme}
\end{figure}

\begin{figure*}
    \includegraphics[width = \textwidth]{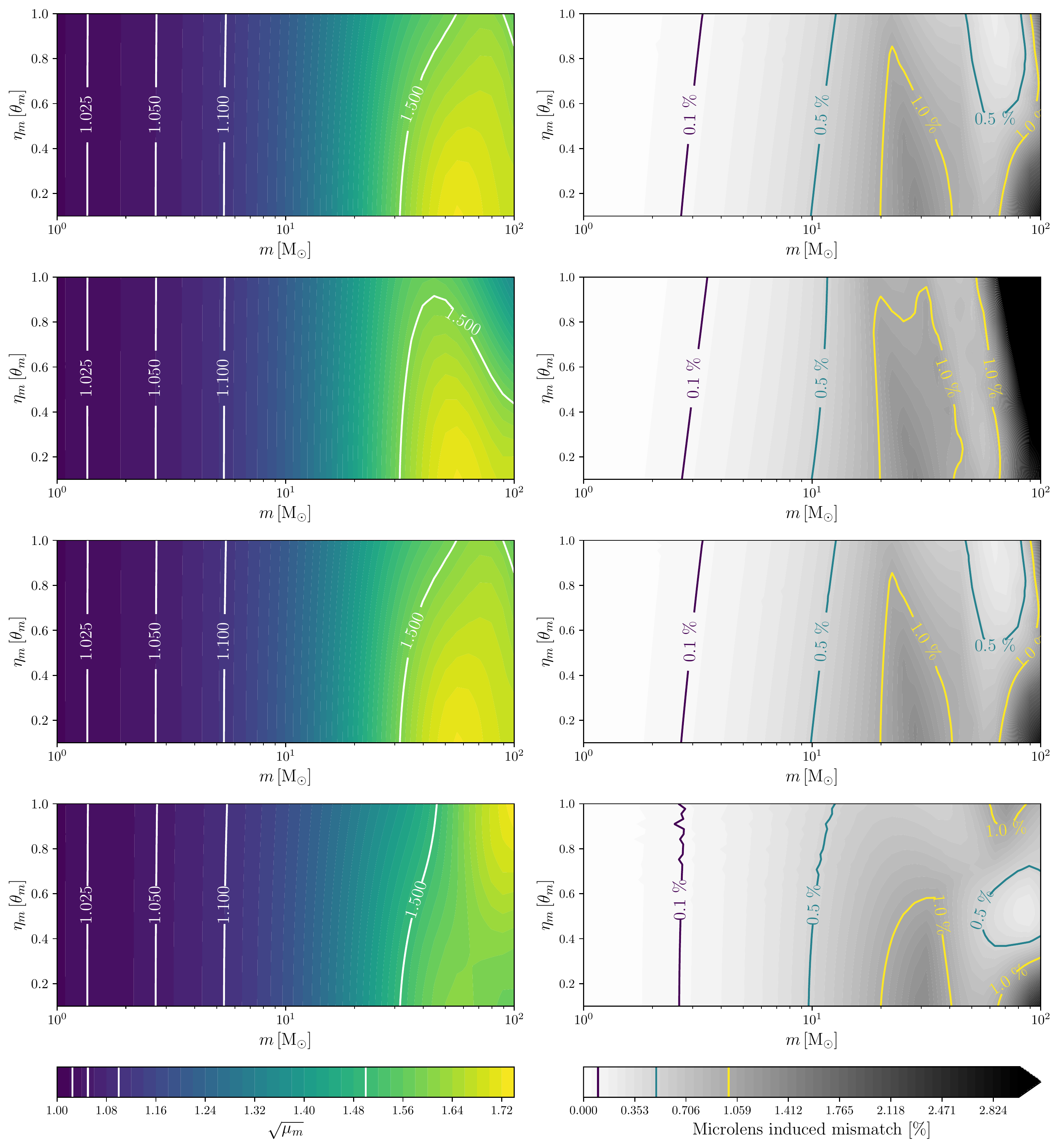}
    \caption{Micro amplification factor $\mu_m$ (left column) and mismatch (right column) for $\phi_m \in \{45\degree, 90\degree, 135\degree, 180\degree\}$ (from top to bottom).
    For $m \lesssim 10 \, \rm M_\odot$, the features are qualitatively similar to those in the case discussed in the main text (see Figures \ref{fig:micromu} and \ref{fig:match}).
    For $m \gtrsim 10$, varying $\phi_m$ does introduce some differences, meaning that the ability to detect microlenses in this higher mass range would at least depend on the three parameters $m$, $\eta_m$, and $\phi_m$}
    \label{fig:angles}
\end{figure*}

In the main text, we only considered the case where the microlenses are placed on the straight line between the SIS lens center and the minimum macroimage. 
Here, with the same macromodel set up, we also vary the direction we place the microlenses from the macroimage, with $\phi_m \in \{45\degree, 90\degree, 135\degree, 180\degree\}$, where $\phi_m$ is the relative angle from the SIS center to the microlens measured at the macroimage as shown in Figure \ref{fig:angle_scheme}.
The case examined in the main text corresponds to $\phi_m = 0\degree$, while the results for the range $180\degree < \phi_m < 360\degree$ are similar to those examined here by symmetry of the model. 
The deviation in micro magnification and the mismatch for the above cases are plotted in Figure \ref{fig:angles}, showing that the microlens displacement relative to the macroimage does not affect the results significantly for $m \lesssim 10 \, \rm M_\odot$.

\bsp	%
\label{lastpage}
\end{document}